%% file: main.tex
\documentclass[11pt,a4paper]{article}

\usepackage[
  top=0.9in, bottom=0.9in, left=0.85in, right=0.85in,
  headheight=28pt, headsep=14pt
]{geometry}

\usepackage[utf8]{inputenc}
\usepackage[T1]{fontenc}
\usepackage{libertine}
\usepackage[varqu,scaled=0.88]{zi4}

\usepackage{deepkernel}

\usepackage{setspace}
\usepackage{fancyhdr}
\usepackage{float}
\usepackage{placeins}

\graphicspath{{figures/}}

\title{%
  \vspace{-1cm}%
  \textbf{\Huge \sys}\\[10pt]
  \textbf{\LARGE The Trust-Native Agentic Operating System}\\[12pt]
  \large\textit{A Paper from the DeepKernel Lab}%
}

\author{%
  Zhenhua Zou\quad
  Sheng Guo\quad
  Qiuyang Zhan\\[3pt]
  Lepeng Zhao\quad
  Shuo Li\quad
  Zhuotao Liu\textsuperscript{\dag}\\[6pt]
  {
  \textsuperscript{\dag}Corresponding author:
  \texttt{zhuotaoliu@tsinghua.edu.cn}}
}

\date{}

\begin{document}

\maketitle
\thispagestyle{fancy}

\input{sections/abstract}

\input{sections/introduction}

\input{sections/background}

\input{sections/principles}

\input{sections/architecture}

\input{sections/analysis}

\input{sections/harness}

\input{sections/discussion}

\input{sections/conclusion}

\bibliographystyle{unsrt}
\bibliography{bibliography/refs}

\end{document}

%% file: sections/abstract.tex
\begin{abstract}

Modern AI agents routinely cross trust boundaries especially in long-horizon tasks: 
they ingest untrusted web and repository content, fuse it with
privileged system instructions, persist intermediate beliefs in
long-term or persistent memory, and invoke privileged tools. 
This pipeline creates a broad surface for malicious payloads to infiltrate through the model's inputs and execute harmful actions via tool calls, raising serious concerns about the level of autonomy we should grant AI agents and also driving a surge in commercial AI agent safeguarding tools.
However, even today's most capable governance stacks remain application-level middleware; 
they co-reside with agents inside a shared process trust boundary.
We argue that what is missing is not another security policy or plugin, but an
\emph{operating-system substrate} that supplies mandatory,
non-bypassable services for identity, input mediation, memory
governance, and execution control.


To systematically close the trust gap, we introduce \sys, the first \emph{trust-native agent operating system} built around a
single premise: security is the first-class design constraint, not an
afterthought bolted onto an LLM runtime.  \sys wraps the agent lifecycle
in a mandatory enforcement boundary organized into four
pillars---Identity, Perception, Cognition, and Execution---each
grounded in classical OS security ideas and lifted to the semantic plane
where autonomous agents actually fail (delegation, prompt injection,
memory poisoning, and tool misuse).

Our central claim is that \emph{structural security is a capability 
multiplier}.  Kernel-managed identity enables trustworthy
cross-organization collaboration; graduated perception replaces brittle
single-point filters; information-flow-controlled memory improves
retrieval fidelity while blocking poisoning; and semantic-to-kernel
execution enforcement lets operators grant broader tool privileges
because the boundary is architecturally non-bypassable.  We position
\sys as the missing OS layer beneath the growing harness
ecosystem---orchestration frameworks, agent runtimes, governance
platforms, and execution sandboxes---and argue through systematic
comparison that \sys is the first integrated agent OS in which security
is the organizing principle across the full lifecycle, enabling stronger
agents rather than merely constraining them.
\end{abstract}

%% file: sections/introduction.tex
\section{Introduction}
\label{sec:intro}

LLM-based AI agents are ceasing to be chat demos and are becoming the \emph{control
plane} through which organizations read, edit, and operate software: a
single stack that couples an LLM to private repositories, terminals,
issue trackers, and---increasingly---other agents.
The first wave that made this legible to practitioners is commercial
\emph{coding-agent} products and IDE-native assistants (e.g.,
Cursor~\cite{anysphere2026cursor}, Claude Code~\cite{anthropic2026claudecode},
and GitHub Copilot/OpenAI Codex-class systems~\cite{github2026copilot,github2026openaicodex})
that already sit inside everyday engineering workflows.
Parallel to those products, high-visibility open-source harnesses such
as OpenClaw~\cite{openclaw2026github,wang2026openclaw} and Nous Research's Hermes agent~\cite{nous2026hermesagent}
have extended the boundary of general AI agents by incorporating diverse message channels for better accessibility and using
self-evolving skills for better personalization.
The research lineage for tool use, delegation, and long-horizon
workflows is well established~\cite{yao2023react,wu2024oscopilot,fourney2024magneticone},
and the library layer has scaled accordingly: representative frameworks
including AutoGPT~\cite{autogpt2025}, LangChain~\cite{langchain2025},
OpenHands~\cite{wang2024openhands}, and MetaGPT~\cite{metagpt2024} now
count hundreds of thousands of GitHub stars in aggregate, while
enterprise-backed stacks such as Microsoft's Agent
Framework~\cite{ms-agent-framework2026}, CrewAI~\cite{crewai2025}, and
Semantic Kernel~\cite{semantickernel2025} underpin production
orchestration.
Mobile and desktop ``computer use'' copilots extend the same pattern to
full GUI surfaces~\cite{ye2025mobileagentv3,zhang2025ufo}.
\emph{Across} products, viral harnesses, and research prototypes, the
architectural constant is unchanged: an LLM fused with multi-source
context and high-privilege tools, without an operating system co-designed to mediate
that fusion end-to-end.

Traditional applications were born into an environment already rich
with OS-level mediation infrastructure: process isolation, virtual memory, file
systems, inter-process communication, and mandatory access control.
These services are so fundamental that no one would attempt to build
a production application without them.  AI agents, by contrast,
operate in an infrastructural vacuum.  They share identity namespaces
(self-declared strings), receive unmediated perception (no reliable
mediation between untrusted content and the LLM context), lack
provenance-aware memory management (no mandatory taint or
information-flow discipline), and execute actions with vague permissions
or unconstrained privileges.  The upshot is not merely a security gap
but an \emph{infrastructure gap}: the same missing substrate that blocks
trustworthy deployment also caps reliability, auditability, and
cross-agent collaboration at scale.

The ecosystem has nevertheless filled up with partial remedies.  We
catalogue them---as we later formalize in \S\ref{sec:harness}---into
four harness tiers that loosely track the lifecycle dimensions of
\S\ref{sec:background} (identity, perception, cognition, execution).
\emph{Orchestration frameworks} such as LangChain/LangGraph~\cite{langchain2025,langgraph2025},
AutoGen~\cite{autogen2024}, CrewAI~\cite{crewai2025}, Microsoft's Agent
Framework~\cite{ms-agent-framework2026}, and Semantic
Kernel~\cite{semantickernel2025} optimize composable workflows and tool
routing.  \emph{Agent runtimes} including AIOS~\cite{mei2024aios},
OpenFang~\cite{openfang2026}, SmythOS SRE~\cite{smythos2026}, and
Letta~\cite{letta2026} combine scheduling and observability with layered
memory (Letta) and, for some stacks, WASM- or cloud-backed resource
isolation (OpenFang, SmythOS).  \emph{Governance platforms}---for
example Microsoft's Agent Governance Toolkit~(AGT)~\cite{microsoft2025governance}
alongside tracing stacks such as LangSmith~\cite{langsmith2025} and
AgentOps~\cite{dong2024agentops}---supply policies, approvals, and audit
pipelines for production agents.  \emph{Execution sandboxes} such as
nono~\cite{nono2026}, E2B~\cite{e2b2026}, and Anthropic's
sandbox-runtime~\cite{anthropic-sandbox2026} isolate OS- or cloud-level
compute from the host.  Each tier advances an important slice of the
stack, yet none is specified as a single mandatory kernel that mediates
every crossing between untrusted input, protected cognition, and
auditable output.

Against classical OS guarantees, today's harness exhibits three
recurring structural weaknesses.  \emph{Fragmentation:} identity
signals, perception filters, memory stores, tool executors, and policy
engines ship as loosely coupled packages with incompatible provenance
labels and no uniform IPC substrate, so compromising or bypassing one
component often voids assumptions elsewhere.  \emph{Blurry trust
boundaries:} even flagship governance stacks remain co-located with
agent business logic---AGT deliberately offers ``application-level
governance, not OS kernel-level isolation,'' with agents and its policy
engine sharing a single process trust boundary~\cite{microsoft2025governance},
so policy, prompts, and tools interleave and neither security nor
application surfaces remain independently auditable or replaceable.
\emph{Non-mandatory enforcement:} defenses appear as optional hooks or
middleware that may inspect tool \emph{parameters} yet cannot bind the
\emph{actual} syscall layout---a child process spawned inside a tool can
still escalate---while sandboxes confine execution without governing how
untrusted perception poisons long-lived memory.

\paragraph{A motivating scenario.}
\label{sec:intro:scenario}
Consider a multi-agent DevOps pipeline: an orchestrator delegates pull
request review and conditional deploy to specialist agents.  A crafted PR
comment carries an indirect prompt-injection payload.  Without
cryptographically grounded delegation, the reviewer cannot authenticate
who authorized the task; without kernel-mediated perception, the payload
reaches the model; without taint-aware memory, the poisoned summary is
archived and later retrieved as trusted context; without
semantic-to-syscall enforcement, a deployer acts on that summary and
pushes malicious code.  Each hop is independently
realistic~\cite{greshake2023indirect,kuntz2025osharm,wang2026openclaw}.
Abstracting from the story, the failure is less a patchwork of unrelated
bugs than the absence of a \emph{composable} mediation narrative: a
traditional OS prevents analogous application compromises because
isolation, mandatory access control, and provenance are always on, span
every resource transition, and compose under one reference monitor.
Agent workloads need the same guarantee lifted to the \emph{semantic}
plane---prompt injections instead of buffer overflows, conversational
context instead of raw address spaces---while still anchoring to the
real syscall surface if enforcement is to be non-bypassable.  Point
solutions reshuffle symptoms; they do not change the end-to-end trust
calculus of a session.  What is missing is therefore not another harness
feature but an \emph{agent operating system}: a mandatory layer that
unifies identity, perception, cognition, and execution mediation.

\paragraph{Proposal and goals.}
We propose \sys, the first trust-native \emph{agent operating system} whose purpose
is to close the structural gap catalogued above: not by accumulating
yet more ad hoc guardrails, but by supplying a \emph{single,
systematically specified, mandatory mediation layer} that is native to
how LLM-based agents actually move trust---through delegation, context
assembly, durable memory, and tool-mediated side effects.
Our objective is practical as much as conceptual: make agent sessions
\emph{auditable}, \emph{composable}, and \emph{deployable at scale}
under realistic environments (richer inputs, broader tools, cross-organization
collaboration), so that
stronger autonomy becomes a reliability and assurance story rather than a
latent incident report.

\paragraph{Design at a glance.}
\sys wraps the agent core in a security kernel that every crossing must
traverse.  \emph{Identity} elevates ``who is acting'' from self-declared
strings to cryptographically grounded enrollment, capability chains, and
mutual attestation so that orchestration, delegation, and agent-to-agent
handoffs carry verifiable provenance.  \emph{Perception} replaces
single brittle filters with a graduated pipeline that treats untrusted
web, repository, and conversational content as labeled inputs whose
risk is reduced \emph{before} it becomes indistinguishable from system
instructions in the model context.  \emph{Cognition} governs what may be
remembered and retrieved: memory is not a passive vector database but an
information-flow--disciplined store in which taint and policy lattice
semantics propagate to individual items, so contaminated beliefs cannot
silently masquerade as operational facts.  \emph{Execution} closes the
last mile from declared tool intent to actual host behavior through a
staged bridge that binds semantic permissions to syscall-level
enforcement (e.g., eBPF hooks and process-tree monitoring), so that
privilege escalation inside a tool implementation cannot outrun the
kernel's view of what was authorized.

\paragraph{Why this is one system, not a toolkit mash-up.}
The four pillars are not interchangeable plugins; they instantiate a
\emph{shared reference-monitor discipline} adapted to the semantic plane.
Classical ideas---non-bypassable mediation, deny-by-default composition,
defense-in-depth, minimal integration surface---are instantiated once,
coherently: identity establishes the principal; perception assigns
initial labels; cognition preserves and propagates those labels across
time; execution enforces the resulting obligations on outward actions.
Returning to the motivating DevOps pipeline (\S\ref{sec:intro:scenario}), the
failure chain is not four unrelated bugs but four missing transitions in
the \emph{same} end-to-end story: impersonation without grounded
identity, injection without kernel-mediated perception, poisoning
without taint-aware cognition, and over-privileged side effects without
semantic-to-syscall execution.  \sys targets that narrative directly:
each pillar supplies an independent invariant so that a compromise in
one layer does not automatically collapse the others, while provenance
and policy flow downward so the system remains a single auditable whole.

\input{figures/ecosystem-stack}

\paragraph{Ecosystem positioning.}
Figure~\ref{fig:ecosystem-stack} situates \sys between agent
applications---commercial coding agents, viral open harnesses, mobile
and desktop ``computer use'' stacks---and the traditional OS kernel.
\sys does not supplant Unix-like kernels; it complements them: the
traditional kernel remains authoritative for address spaces, devices,
and low-level isolation, while \sys becomes authoritative for \emph{agent
semantics}---who may speak for whom, which content may enter the LLM
context, how memory items inherit trust, and which tool behaviors are
actually permitted on the host.  Together, the two layers form what we
call an \emph{AI-Native OS}: semantic mediation with a syscall backstop.
Host kernels already enforce isolation and mandatory access over
kernel-addressable objects---files, processes, sockets---yet the residual
problem is not another layer of labels on those objects.  Agents fuse
untrusted natural language with privileged automation across sessions,
organizations, and long horizons, so the crossings that must be mediated
include intents, memories, and plans that kernels neither model nor
observe.  \sys supplies the missing mandatory services on the semantic
plane while anchoring outward effects to the traditional kernel where they
ultimately become syscalls.  We borrow from the MAC era only its
\emph{architectural shape}---always-on, non-bypassable mediation that
composes into a deployable trust story---rather than duplicating kernel
object enforcement.  Treating structural security this way makes it a
\emph{capability multiplier}: operators can justify richer untrusted
inputs, broader tool privileges, and cross-domain collaboration because
the boundary is non-bypassable by construction.  The remainder of this
section summarizes our contributions.

\paragraph{Contributions.}
This paper contributes the following, in service of the positioning
above:
\begin{enumerate}
  \item A \textbf{unified lifecycle diagnosis} that reframes today's agent
    stacks around four trust-critical phases---identity, perception,
    cognition, and execution---and analyzes what shape of OS-level services the 
    model AI agents need and why traditional low-level OS primitives fall short.
    We go deeper to reveal the security and capabilities consequences of the lack of 
    such semantic-layer OS mediations, motivating our insight that
    an OS-shaped substrate is prerequisite rather than optional
    (\S\ref{sec:background}).
  \item The \textbf{\sys architecture and design principles}: a mandatory
    security kernel organized into the four pillars above, guided by
    five principles distilled from classical OS security (external
    reference monitor, deny-by-default policy intersection, lifecycle
    defense-in-depth, semantic mediation with syscall backstop, and
    a narrow agent--kernel integration boundary to model, tool, and
    storage capability).  We present the two trust anchors (a
    remote registry for enrollment and credentials, and a local agent
    kernel), the three integration adapters through which frameworks
    attach, and the end-to-end data and control paths that keep mediation
    outside the manipulable LLM context (\S\ref{sec:principles},
    \S\ref{sec:architecture}).
  \item A \textbf{structured security analysis} that states explicit
    properties per pillar, articulates how cryptographic, information-flow,
    and kernel-backed guarantees compose, and clarifies the residual
    threat model---including what remains when policies are mis-set and
    how the semantic--syscall bridge is intended to behave under stress
    (\S\ref{sec:analysis}).
  \item A \textbf{scientific positioning} of \sys within the rapidly
    evolving harness landscape: a tiered map of orchestration frameworks,
    agent runtimes, governance platforms, and execution sandboxes,
    followed by a systematic comparison to representative systems (e.g.,
    Microsoft's Agent Governance Toolkit, AIOS-class runtimes, and
    sandbox offerings).  The goal is not to claim feature parity but to
    show where mandatory lifecycle mediation is absent today and how an
    OS-first layer complements capability-first and policy-first stacks
    (\S\ref{sec:harness}).
\end{enumerate}

\paragraph{Roadmap.}
\S\ref{sec:background} develops the lifecycle model and threat surface.
\S\ref{sec:principles} and \S\ref{sec:architecture} present the design
principles and architectural instantiation of \sys.
\S\ref{sec:analysis} analyzes security properties and composition.
\S\ref{sec:harness} maps the broader ecosystem, compares against related systems, and argues for the OS
layer's role in the trusted computing base.
\S\ref{sec:discussion} discusses limitations and research directions, and
\S\ref{sec:conclusion} concludes.

%% file: figures/ecosystem-stack.tex
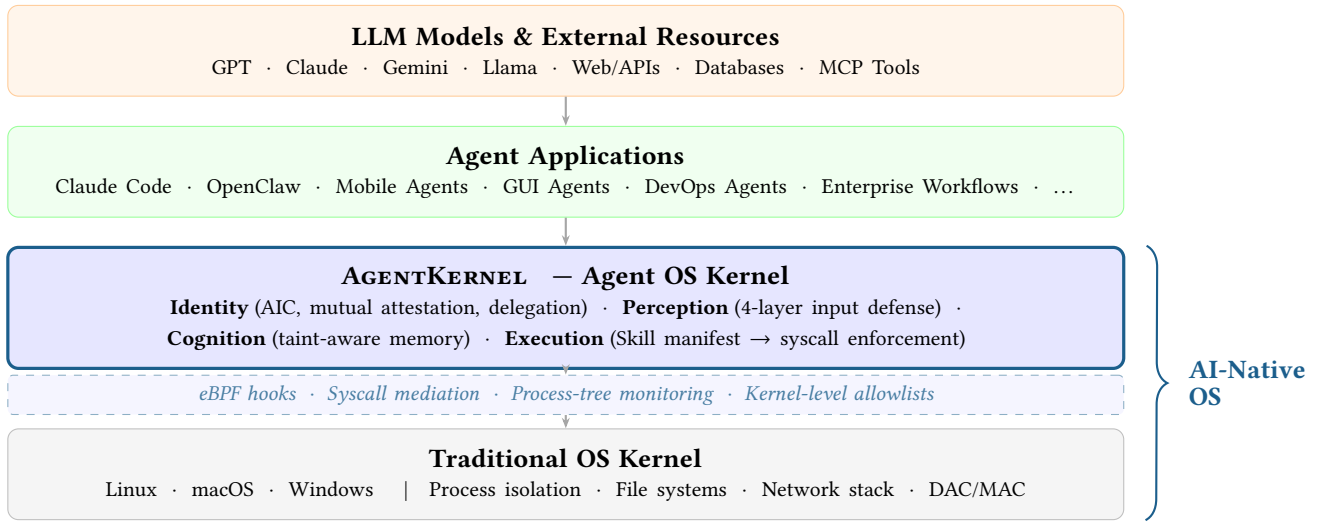
\begin{figure*}[t]
  \centering
  \begin{tikzpicture}[
    layer/.style={
      draw, rounded corners=4pt, minimum height=1.2cm,
      text width=14.5cm, align=center, font=\small
    },
    arr/.style={-{Stealth[length=4pt]}, thick, gray!70},
    brace/.style={decorate, decoration={brace, amplitude=8pt, mirror}, thick, dkblue},
    lbl/.style={font=\footnotesize\itshape, text=gray!80}
  ]

    \node[layer, fill=orange!8, draw=orange!40]
      (llm) at (0, 4.8)
      {\textbf{LLM Models \& External Resources}\\[-1pt]
       \scriptsize GPT \;\textperiodcentered\; Claude
       \;\textperiodcentered\; Gemini \;\textperiodcentered\;
       Llama \;\textperiodcentered\; Web/APIs
       \;\textperiodcentered\; Databases \;\textperiodcentered\;
       MCP Tools};

    \node[layer, fill=green!6, draw=green!40]
      (agents) at (0, 3.2)
      {\textbf{Agent Applications}\\[-1pt]
       \scriptsize Claude Code \;\textperiodcentered\;
       OpenClaw \;\textperiodcentered\;
       Mobile Agents \;\textperiodcentered\;
       GUI Agents \;\textperiodcentered\;
       DevOps Agents \;\textperiodcentered\;
       Enterprise Workflows \;\textperiodcentered\;
       \ldots};

    \node[layer, fill=blue!10, draw=dkblue, line width=1.2pt,
          minimum height=1.6cm]
      (dk) at (0, 1.4)
      {\textbf{\sys\; --- Agent OS Kernel}\\[-1pt]
       \scriptsize
       \textbf{Identity}\;(AIC, mutual attestation, delegation)
       \;\textperiodcentered\;
       \textbf{Perception}\;(4-layer input defense)
       \;\textperiodcentered\;
       \textbf{Cognition}\;(taint-aware memory)
       \;\textperiodcentered\;
       \textbf{Execution}\;(Skill manifest $\to$ syscall enforcement)};

    \node[draw=dkblue!50, dashed, fill=blue!4,
          rounded corners=2pt, minimum height=0.5cm,
          text width=14.5cm, align=center, font=\scriptsize\itshape,
          text=dkblue!80]
      (overlap) at (0, 0.25)
      {eBPF hooks \;\textperiodcentered\;
       Syscall mediation \;\textperiodcentered\;
       Process-tree monitoring \;\textperiodcentered\;
       Kernel-level allowlists};

    \node[layer, fill=gray!8, draw=gray!50]
      (os) at (0, -0.8)
      {\textbf{Traditional OS Kernel}\\[-1pt]
       \scriptsize Linux \;\textperiodcentered\;
       macOS \;\textperiodcentered\; Windows
       \quad\textbar\quad
       Process isolation \;\textperiodcentered\;
       File systems \;\textperiodcentered\;
       Network stack \;\textperiodcentered\;
       DAC/MAC};

    \draw[arr] (llm.south) -- (agents.north);
    \draw[arr] (agents.south) -- (dk.north);
    \draw[arr] (dk.south) -- (overlap.north);
    \draw[arr] (overlap.south) -- (os.north);

    \draw[brace]
      ([xshift=8pt]os.south east) --
      ([xshift=8pt]dk.north east)
      node[midway, right=12pt, align=left, font=\small\bfseries,
           text=dkblue]
      {AI-Native\\[-2pt] OS};

  \end{tikzpicture}
  \caption{%
    \sys's position in the AI agent technology stack.
    Agent applications (Claude Code, OpenClaw, mobile/GUI agents, etc.)
    sit atop \sys, which mediates all interactions with LLM models,
    tools, and the outside world.
    \sys and the traditional OS kernel jointly form the
    \emph{AI-Native OS}: \sys provides semantic-level services
    (identity, perception, cognition, execution governance) while
    leveraging the traditional kernel for syscall-level enforcement
    via eBPF hooks and process-tree monitoring.%
  }
  \label{fig:ecosystem-stack}
\end{figure*}

%% file: sections/background.tex
\section{Background and Motivation}
\label{sec:background}

\subsection{The Agent Lifecycle}
\label{sec:bg:lifecycle}

A modern LLM-powered agent operates through a continuous loop of
\emph{perception}, \emph{cognition}, and \emph{execution}.  During
\textbf{perception}, the agent ingests inputs from diverse
sources---user messages, tool outputs, web pages, API responses, and
messages from other agents.  During \textbf{cognition}, it reasons over
these inputs, retrieves and updates long-term memory, formulates plans,
and decomposes goals into sub-tasks.  During \textbf{execution}, it
invokes tools, calls APIs, writes files, sends messages, or delegates
sub-tasks to other agents.  Cutting across all three phases is
\textbf{identity}: the agent must establish \emph{who} it is, \emph{what}
it is permitted to do, and \emph{whom} it is interacting with.

\begin{figure}[t]
  \centering
  \begin{tikzpicture}[
    node distance=1.2cm and 2.8cm,
    box/.style={draw, rounded corners=4pt, minimum width=3cm,
                minimum height=1.1cm, align=center, font=\small},
    arrow/.style={-{Stealth[length=5pt]}, thick},
    lbl/.style={font=\scriptsize\itshape, text=gray, above=3pt}
  ]
    \node[box, fill=red!8]   (input)  {Input World\\[-2pt]\footnotesize(untrusted)};
    \node[box, fill=blue!8,  right=of input]  (kernel) {\sys Security\\[-2pt]\footnotesize Kernel};
    \node[box, fill=green!8, right=of kernel] (output) {Output World\\[-2pt]\footnotesize(auditable)};

    \draw[arrow] (input)  -- node[lbl, midway] {normalize, tag, filter} (kernel);
    \draw[arrow] (kernel) -- node[lbl, midway] {verify, enforce, audit} (output);

    \node[box, fill=blue!3, below=0.9cm of kernel, minimum width=3cm]
      (core) {Protected\\Agent Core};
    \draw[arrow, <->] (kernel) -- (core);
  \end{tikzpicture}
  \caption{Three-domain model.  All inputs traverse the \sys kernel
    before reaching the protected agent core; all outputs are
    identity-verified, intent-aligned, and permission-converged before
    leaving the kernel.}
  \label{fig:three-domains}
\end{figure}
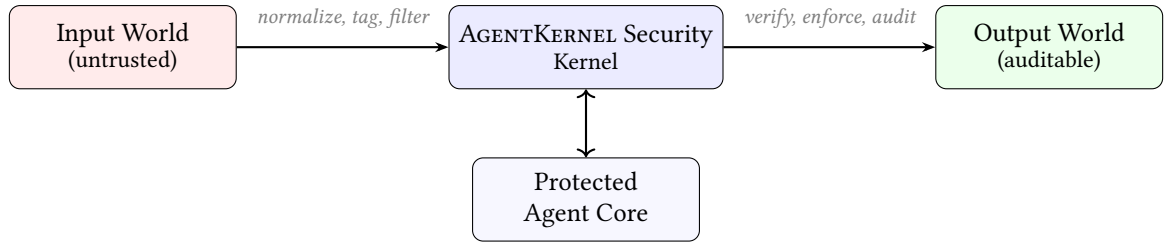

We conceptualize the agent's operating environment as three trust
domains, illustrated in Figure~\ref{fig:three-domains}:

\begin{itemize}
  \item \textbf{Input World} (untrusted by default): user instructions,
    external agents, tool outputs, web content, system events.
  \item \textbf{Agent Core} (protected): reasoning engine, context and
    memory stores, session and trajectory state.
  \item \textbf{Output World} (auditable, least-privilege): tool
    execution, data sinks, agent-to-agent delegation, audit logs.
\end{itemize}

Every transition between domains represents a \emph{trust boundary
crossing} that requires mediation.
Yet in most contemporary agent
frameworks, these crossings are unguarded.

\subsection{What Agents Need from an Operating System, and Why Classical Primitives Fall Short}
\label{sec:bg:os-needs}

Traditional applications benefit from four categories of OS services:
\emph{identity and access control} (UIDs, file permissions, SELinux
labels), \emph{input mediation} (device drivers, protocol parsers),
\emph{memory management} (virtual memory, page tables, copy-on-write),
and \emph{execution governance} (process isolation, capabilities,
sandboxing).  By analogy, AI agents require semantic-level counterparts
to these services:

\begin{enumerate}
  \item \textbf{Identity management.}  Agents need cryptographic
    identity that binds who built the agent, what code it runs, and who
    authorized its deployment---enabling trustworthy A2A collaboration
    and accountable delegation chains.
  \item \textbf{Input mediation.}  Content from untrusted sources must
    be normalized, tagged, and filtered \emph{before} entering the LLM
    context---not as an optional guardrail, but as a mandatory OS
    service analogous to a device driver sanitizing hardware input
    (cf.\ Figure~\ref{fig:three-domains}).
  \item \textbf{Memory management with provenance.}  As agents gain
    persistent memory, they need information-flow control that tracks
    the origin and trustworthiness of each memory item---enabling better
    retrieval decisions and preventing persistent contamination.
  \item \textbf{Execution governance.}  Agent actions must be mediated
    from semantic intent down to actual system calls, ensuring that
    declared permissions match runtime behavior---enabling operators to
    safely grant broader tool access with confidence that the enforcement
    boundary holds.
\end{enumerate}

Interpreting the lifecycle of~\S\ref{sec:bg:lifecycle}, these four
counterpart services respectively target \emph{Identity}, inputs at the
\emph{Perception} boundary, provenance-aware \emph{Cognition} state, and
policy-convergent \emph{Execution}---the same dimensions used in
Table~\ref{tab:threats}---and correspond to the mediated trust-boundary
crossings in Figure~\ref{fig:three-domains}.

Traditional operating systems nevertheless provide powerful primitives at
a lower layer of abstraction:
discretionary and mandatory access control, process isolation,
sandboxing, and network filtering.  These mechanisms operate at the
\emph{syscall level}---they mediate access to files, sockets, and
devices.  However, agent-specific gaps operate at the
\emph{semantic level}:

\begin{itemize}
  \item An agent may present valid credentials, file descriptors, or IPC
    handles while syscall policies remain blind to cryptographic binding
    between running code, its builder, and the authorizing principal for
    a delegation chain.
  \item A prompt injection embedded in a tool's JSON response is
    well-formed data---it violates no system call policy.
  \item An agent ``legitimately'' calling a file-write API but with
    content derived from a poisoned memory entry is invisible to
    syscall-level checks.
  \item An agent delegating to a sub-agent with escalated capabilities is
    a normal IPC operation from the OS perspective.
\end{itemize}

The gap between OS-level services and agent-level services is analogous
to the gap that motivated application-layer firewalls: just as TCP/IP
packet filtering cannot detect SQL injection, syscall mediation cannot
detect prompt injection (we use this comparison only as intuition, not as
a formal equivalence claim).  What is needed is a new layer of
\emph{semantic-level OS services} that understands agent intents,
memory provenance, and capability boundaries---while still leveraging
OS-level enforcement as the ultimate backstop.

\subsection{Infrastructure Gaps and Their Consequences}
\label{sec:bg:threats}

Absent or incomplete semantic-layer services couple security breaches with
artificial capability ceilings.
Table~\ref{tab:threats} summarizes representative paired consequences along
each lifecycle dimension.

\begin{table}[t]
  \centering
  \caption{Infrastructure gaps across the agent lifecycle: missing OS
    services create both security vulnerabilities and capability
    limitations.}
  \label{tab:threats}
  \small
  \begin{tabularx}{\linewidth}{@{}lXX@{}}
    \toprule
    \textbf{Dimension} & \textbf{Security Consequence} & \textbf{Capability Consequence} \\
    \midrule
    \multirow{2}{*}{Identity}
      & Impersonation and unauthorized multi-hop delegation without binding principals to code and authorization
      & Verifiable delegation chains across organizations stall; operators fall back on pairwise trust or manual approval gates \\
      & Unverifiable third-party skills/plugins enabling supply-chain capability injection
      & Version freezes and strict whitelists block safe participation in open third-party capability ecosystems \\
    \midrule
    \multirow{2}{*}{Perception}
      & Prompt injection via tools, web pages, and files; untrusted content co-mingled with privileged instructions
      & Browser- and repository-backed toolchains are avoided or over-stripped, narrowing tasks an agent can safely cover \\
      & Encoding bypasses, jailbreaks, and spoofed UI or sensor readings that misstate device or application state
      & Multimodal and high-entropy pipelines require human checkpoints, capping throughput and end-to-end autonomy \\
    \midrule
    \multirow{2}{*}{Cognition}
      & Cross-session memory poisoning; instruction--data confusion in long, multi-source context windows
      & Cross-session RAG and long-horizon memory cannot scale without per-entry provenance or costly write-side audits \\
      & Cross-tenant memory contamination; weak edge-side summarisation or routing exploited as a chain weak link
      & Shared memory pools across agents or business units are blocked; teams revert to siloed stores and manual reconciliation \\
    \midrule
    \multirow{2}{*}{Execution}
      & Narratives that diverge from ground-truth tool traces; irreversible actions without compensating controls
      & Regulated and embodied deployments lack high-assurance automation without signed, replayable execution artefacts \\
      & Privilege creep and malicious-by-construction tools composed into high-impact effects
      & Dynamic tool composition is disallowed or confined to narrow sandboxes, preventing end-to-end agentic workflows \\
    \bottomrule
  \end{tabularx}
\end{table}

\paragraph{Identity gaps.}
Most frameworks treat identity as self-declared names or API keys, not as
cryptographic evidence binding builders, binaries or prompts, and
authorized capability sets~\cite{south2025identitymanagement,wang2026openclaw}.
That gap enables impersonation and \emph{supply-chain} injection of
third-party skills, plugins, and tools whose provenance cannot be
checked~\cite{yang2025agentprotocols}.
In GUI- and multimodal-driven deployments, presentation-layer spoofing
(package names, icons, screenshots) further blurs ``which application or
principal the agent is acting for,'' coupling identity failures to
downstream perception and execution errors.

\paragraph{Perception gaps.}
Direct and indirect prompt injection remains the dominant
vector~\cite{greshake2023indirect}: without mandatory mediation, tool
outputs, web pages, files, and agent-to-agent messages are ingested as
unlabeled high-entropy text indistinguishable from operator instructions.
\emph{State deception}---transparent overlays, adversarial screenshots, or
noisy sensor streams---poisons the agent's world model even when syscall
policies are satisfied, a failure mode highlighted by safety analyses of
computer-use agents~\cite{kuntz2025osharm}.
Operationally, teams respond by stripping multimodal context or inserting
human-in-the-loop review, which caps automation even when no attacker is
present.

\paragraph{Cognition gaps.}
Persistent stores let a single poisoned entry influence later
sessions~\cite{li2025memos,kang2025memoryos}, analogous to a TOCTOU class of
bug where the unsafe act (retrieval) is separated in time from the
poisoning write.
Long contexts assembled from heterogeneous sources exacerbate
instruction--data confusion, biasing planning and tool selection without
surfacing low-confidence branches to users.
Heterogeneous deployments that route summarisation, filtering, or
delegation through resource-constrained edge components further concentrate
risk on the weakest semantic link in the chain~\cite{wei2025agentxpu}.

\paragraph{Execution gaps.}
Agents often hold broad, slowly changing tool permissions while tasks demand
fine-grained, ephemeral authority; bridging that gap without telemetry
invites \emph{privilege creep} and composition of nominally benign tools
into high-impact sequences.
Malicious or trojaned skills and scripts remain dangerous even when the LLM
``behaves,'' because execution is ultimately local code with ambient access.
The \emph{plan--trace alignment} problem---natural-language rationales that
drift from signed execution records---collapses auditability for regulated
and irreversible (e.g., financial or physical) actions; recent systems
mitigate this class of failure by cross-checking claims against signed tool
receipts~\cite{basu2026toolreceipts}.

Collectively, these gaps call for a \emph{single semantic mediation layer}
that is mandatory for the agent lifecycle rather than bolted on per
integration point.  Section~\ref{sec:principles} states the design
principles we adopt for such a layer, and Section~\ref{sec:architecture}
presents \sys as a concrete architecture that instantiates them.

%% file: sections/principles.tex
\section{Design Principles}
\label{sec:principles}

Before presenting \sys's architecture, we distill five design
principles that guide every aspect of the system.  These principles
are not merely implementation guidelines; they constitute the
intellectual core of our position and distinguish \sys from
capability-first agent OS proposals.

\begin{principle}[Security as an Independent \& Mandatory Trusted Layer]
The agent security kernel must operate as an \emph{external,
non-bypassable reference monitor}---independent of the agent's own
reasoning loop.
\end{principle}

In classical OS security, the reference monitor
concept~\cite{anderson1972planning} requires that every access be
mediated by a tamper-proof authority that is itself not subject to the
policy it enforces.  The same principle applies to agents, but with a
critical twist: the agent's ``reasoning loop'' (the LLM) is inherently
manipulable through prompt injection.  Embedding security logic
\emph{inside} the LLM's context---as many guardrail approaches
do---means that a successful injection can simultaneously subvert the
security check and the protected action.  \sys therefore places its
security kernel \emph{outside} the LLM context, mediating all inputs and
outputs through an independent enforcement layer.  The agent cannot
bypass the kernel because the kernel controls the only interfaces
(adapters) through which the agent can reach the LLM, tools, and
storage.

\begin{principle}[Deny-by-Default with Policy Intersection]
When multiple policies apply, the effective permission set is the
\emph{intersection}---never the union---of all applicable policy sets.
\end{principle}

In multi-stakeholder agent deployments, permissions originate from
several sources: the developer declares a maximum capability set at build
time, the operator grants a runtime subset, and the agent security kernel
further constrains based on context.  If the combination rule were a
\emph{union}, any single permissive policy could override all others.
\sys mandates \emph{intersection}: the effective permission is always
\[
  \mathcal{S}_{\text{eff}} \;=\;
    \mathcal{S}_{\text{dev}} \;\cap\;
    \mathcal{S}_{\text{ops}} \;\cap\;
    \mathcal{S}_{\text{ctx}}
\]
This ``converge-on-strictest'' rule applies at every level: AIC
capability chains (\S\ref{sec:arch:identity}), skill permission
composition (\S\ref{sec:arch:execution}), and cross-agent session tokens
(\S\ref{sec:arch:identity}).

\begin{principle}[Defense-in-Depth Across the Agent Lifecycle]
Each of the four pillars (Identity, Perception, Cognition, Execution)
provides \emph{independent, composable} defense layers; compromise of
one pillar does not cascade to others.
\end{principle}

Defense-in-depth is a well-established security doctrine, yet most agent
security approaches implement it \emph{within} a single layer (e.g.,
multiple prompt-injection filters in sequence).  \sys extends
defense-in-depth \emph{across the entire agent lifecycle}: identity
prevents impersonation regardless of whether perception catches an
injection; perception sanitizes inputs regardless of whether cognition
tracks taint; cognition isolates contaminated memory regardless of
whether execution blocks the resulting action.  Each pillar reduces the
residual risk left by the others, and their independence means that a
zero-day in one does not unravel the entire defense.
Figure~\ref{fig:defense-depth} visualizes this cross-lifecycle defense
architecture.

\begin{figure*}[t]
  \centering
  \input{figures/defense-depth}
  \caption{%
    Defense-in-depth across the agent lifecycle.  Each pillar
    independently intercepts a distinct threat class.  Provenance
    labels flow across pillars (dashed arrows), enabling downstream
    pillars to make informed decisions based on upstream assessments.
    Compromise of one pillar does not cascade to others.%
  }
  \label{fig:defense-depth}
\end{figure*}

\begin{principle}[Semantic-Level Mediation]
The agent security kernel mediates at the \emph{semantic level}
(intents, plans, memory entries) while also leveraging OS-level
enforcement (eBPF, process sandboxes) as the ultimate backstop.
\end{principle}

Traditional OS security operates at the syscall boundary---it can
prevent a process from opening a file but cannot determine whether the
file's \emph{content} was derived from a poisoned memory entry.  Agent
security requires a higher-level mediation point that understands
\emph{what the agent intends} and \emph{why}, not just \emph{which
syscall} it issues.  \sys achieves this through a dual-layer approach:
semantic mediation at the tool-call, memory-access, and plan-evaluation
levels (\S\ref{sec:arch:perception}--\ref{sec:arch:execution}),
backstopped by kernel-level enforcement via eBPF hooks that constrain
the actual system calls (\S\ref{sec:arch:execution}).  This dual-layer
design closes the semantic--syscall gap identified in
\S\ref{sec:bg:os-needs}.

\begin{principle}[Narrow Agent--Kernel Integration Boundary]
Relative to the manipulable agent loop, \sys exposes exactly three narrow
integration interfaces---the LLM, tool, and storage adapters---bounding
the \emph{agent-visible} surfaces of inference, tool execution, and durable
memory (distinct from the kernel's internal modularization).
\end{principle}

The four pillars (Identity, Perception, Cognition, Execution) specify
\emph{how} policy is enforced across the lifecycle (Principle~3;
Figure~\ref{fig:defense-depth}): composable invariants in the trusted
enforcement layer.  The three adapters specify \emph{where} untrusted agent
and framework code may reach model inference, tools, and durable memory.
Pillar services attach along these paths (e.g., perception on inbound
model-bound content); identity is a kernel-managed control-plane substrate,
not a fourth application-side ``identity adapter.''  Restricting
co-evolving integration glue to three choke points instantiates \emph{economy
of mechanism} and \emph{complete mediation} for that resource model.  This
is \emph{not} the claim that ``three APIs'' implement the pillars, nor that
the TCB is small: GAR, the Agent Kernel, and pillar logic remain inside the
TCB (\S\ref{sec:architecture}).

Like Principle~1, soundness requires \emph{non-bypass}: the agent and its
orchestration fabric must not retain \emph{parallel} access to the same
model endpoints, tool runtimes, or stores outside \sys{} (ambient credentials, direct
tool transports, filesystem/DB paths that bypass the storage adapter, \emph{etc.}).
Otherwise mediation is incomplete regardless of adapter narrowness.
Deployment shape is secondary---embedded library, gateway, sidecar, or
tool facade---provided the mandatory trusted layer remains the sole authority
over those resources.

The adapters then yield three integration properties:
\begin{itemize}
  \item \textbf{Scoped complete mediation.}  All policy-relevant traffic
    for the three capability classes traverses the adapters, giving one
    auditable seam to harden or replace glue code.
  \item \textbf{Kernel--application decoupling.}  The kernel can be
    audited, tested, or upgraded without rewriting agent logic; frameworks
    can be swapped without re-implementing pillar semantics.
  \item \textbf{Explicit evolution rule.}  A new externally visible
    capability class beyond model, tool, and storage effects warrants a
    new integration abstraction, not ad hoc bypasses.
\end{itemize}

\parab{Relationship to classical OS principles}
These five principles are deeply rooted in classical OS security:
Principle~1 extends the reference
monitor~\cite{anderson1972planning}; Principle~2
generalizes the principle of least privilege and
fail-safe defaults~\cite{saltzer1975protection};
Principle~3 adapts defense-in-depth to the agent
lifecycle; Principle~4 bridges the semantic--syscall
gap; and Principle~5 applies economy of mechanism and complete
mediation~\cite{saltzer1975protection} at the agent--kernel integration
boundary.  What is new is their
\emph{application to autonomous AI agents}---a domain where the
``user'' is an LLM, the ``application'' is an agentic loop, and the
``data'' may itself contain executable intent.

%% file: figures/defense-depth.tex
\begin{tikzpicture}[
  node distance=0.5cm and 1.8cm,
  pillar/.style={
    draw=dkborder, rounded corners=4pt, minimum width=2.6cm,
    minimum height=2.2cm, align=center, font=\small, fill=white
  },
  flow/.style={-{Stealth[length=5pt]}, thick, dashed, color=dkaccent},
  defend/.style={-{Stealth[length=5pt]}, thick, color=dkred},
  lbl/.style={font=\scriptsize\sffamily, text=dkdarkgray},
  risk/.style={font=\scriptsize\itshape, text=dkred},
]

\node[pillar, fill=dkorange!8] (id) at (0,0) {
  \textbf{Identity}\\[2pt]
  {\scriptsize Crypto AIC}\\
  {\scriptsize 4-dim binding}\\
  {\scriptsize Delegation chains}
};

\node[pillar, fill=yellow!8, right=of id] (perc) {
  \textbf{Perception}\\[2pt]
  {\scriptsize P1 Source tags}\\
  {\scriptsize P2 Rule filter}\\
  {\scriptsize P3 Semantic FW}\\
  {\scriptsize P4 Jailbreak det.}
};

\node[pillar, fill=cyan!8, right=of perc] (cog) {
  \textbf{Cognition}\\[2pt]
  {\scriptsize Lattice taint}\\
  {\scriptsize Item-level labels}\\
  {\scriptsize Provenance chains}
};

\node[pillar, fill=dkgreen!8, right=of cog] (exec) {
  \textbf{Execution}\\[2pt]
  {\scriptsize E1 Policy rules}\\
  {\scriptsize E2 LLM validate}\\
  {\scriptsize E3 eBPF hooks}\\
  {\scriptsize E4 Plan--trace}
};

\draw[flow] (id.east) -- (perc.west)
  node[above, midway, lbl] {trust levels};
\draw[flow] (perc.east) -- (cog.west)
  node[above, midway, lbl] {provenance};
\draw[flow] (cog.east) -- (exec.west)
  node[above, midway, lbl] {taint labels};

\node[risk, above=0.7cm of id] (t1) {Spoofing};
\node[risk, above=0.7cm of perc] (t2) {Injection};
\node[risk, above=0.7cm of cog] (t3) {Poisoning};
\node[risk, above=0.7cm of exec] (t4) {Escalation};

\draw[defend] (t1) -- (id);
\draw[defend] (t2) -- (perc);
\draw[defend] (t3) -- (cog);
\draw[defend] (t4) -- (exec);

\node[font=\sffamily\small\itshape, text=dkdarkgray, anchor=north]
  at ($(perc.south)!0.5!(cog.south) + (0,-0.6)$)
  {Each pillar independently reduces residual risk;
   compromise of one does not cascade to others.};

\end{tikzpicture}

%% file: sections/architecture.tex
\section{The \sys Architecture}
\label{sec:architecture}

This section presents the \sys architecture at the conceptual and
structural level.  We begin with the agent-kernel contract and a
structural overview grounded in the three-domain model of
Section~\ref{sec:background}, then detail each of the four security
pillars.

\subsection{Architectural Overview}
\label{sec:arch:overview}

Traditional OS kernels mediate access to physical resources through a
mandatory, non-bypassable reference monitor.  Agents, however, fail at
the \emph{semantic} boundary: untrusted content, poisoned memory, and
misattributed actions cross into reasoning and tool execution without a
single choke point that can enforce end-to-end policy.  We therefore
treat an \emph{agent OS kernel} as a semantic reference monitor: every
interaction with the outside world must pass through kernel-enforced
services that compose into a coherent trust argument.

Concretely, such a kernel must supply four integrated services, each
addressing a distinct class of failures that Section~\ref{sec:background}
isolates across the agent lifecycle:

\begin{enumerate}
  \item \textbf{Cryptographic agent identity} binds builders, the code or
    configuration the agent actually runs, and authorization to
    deploy---mitigating impersonation and supply-chain substitution while
    enabling verifiable agent-to-agent collaboration.  Identity material
    (signing and attestation keys) is held and used only through the
    kernel, never as an ambient secret in application space.
  \item \textbf{Graduated input mediation} applies multiple independent
    filters before content reaches the LLM context, trading off latency for
    defense in depth against injection, spoofing, and high-entropy
    confusion---improving safety without forcing operators to discard
    entire modalities.
  \item \textbf{Information-flow control} over agent memory tracks
    provenance and taint at per-entry granularity so retrieval cannot
    silently amplify stale or adversarial state---supporting both poisoning
    resistance and provenance-aware cognition.
  \item \textbf{Execution governance} closes the loop from stated intent
    through tool plans to syscall-level effects, checking alignment with
    operator policy and producing replayable evidence when actions are
    sensitive or irreversible---restoring auditability while still allowing
    dynamic tool expansion under explicit policy.
\end{enumerate}

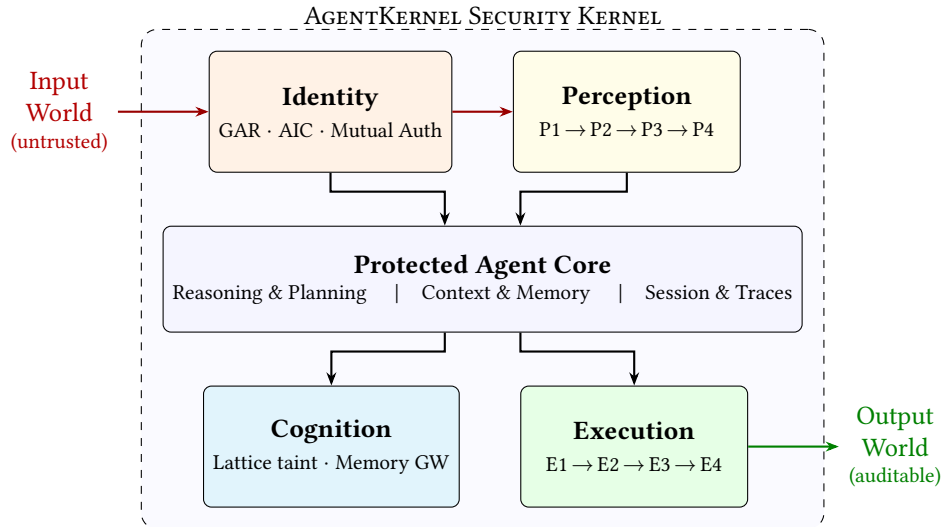
\begin{figure*}[t]
  \centering
  \begin{tikzpicture}[
    node distance=0.6cm and 0.8cm,
    phase/.style={draw, rounded corners=3pt, minimum width=3cm,
                  minimum height=1.6cm, align=center, font=\small},
    core/.style={draw, rounded corners=3pt, fill=blue!4,
                 minimum width=6.8cm, minimum height=1.4cm,
                 align=center, font=\small},
    zone/.style={draw, dashed, rounded corners=6pt, inner sep=8pt},
    arrow/.style={-{Stealth[length=5pt]}, thick},
    lbl/.style={font=\scriptsize\itshape, text=gray}
  ]
    \node[phase, fill=orange!10] (identity)
      {\textbf{Identity}\\{\scriptsize GAR $\cdot$ AIC $\cdot$ Mutual Auth}};
    \node[phase, fill=yellow!10, right=of identity] (perception)
      {\textbf{Perception}\\{\scriptsize P1\,\textrightarrow\,P2\,\textrightarrow\,P3\,\textrightarrow\,P4}};

    \node[core, below=0.7cm of $(identity.south east)!0.5!(perception.south west)$]
      (agentcore) {\textbf{Protected Agent Core}\\[-2pt]
      {\scriptsize Reasoning \& Planning \quad$\vert$\quad
       Context \& Memory \quad$\vert$\quad Session \& Traces}};

    \node[phase, fill=cyan!10, below=0.7cm of agentcore, xshift=-2cm]
      (cognition)
      {\textbf{Cognition}\\{\scriptsize Lattice taint $\cdot$ Memory GW}};
    \node[phase, fill=green!10, right=of cognition]
      (execution)
      {\textbf{Execution}\\{\scriptsize E1\,\textrightarrow\,E2\,\textrightarrow\,E3\,\textrightarrow\,E4}};

    \draw[arrow] (identity.south) -- ++(0,-0.25) -| ([xshift=-0.5cm]agentcore.north);
    \draw[arrow] (perception.south) -- ++(0,-0.25) -| ([xshift=0.5cm]agentcore.north);
    \draw[arrow] ([xshift=-0.5cm]agentcore.south) -- ++(0,-0.25) -| (cognition.north);
    \draw[arrow] ([xshift=0.5cm]agentcore.south) -- ++(0,-0.25) -| (execution.north);

    \begin{scope}[on background layer]
      \node[zone, fill=blue!2,
            fit=(identity)(perception)(agentcore)(cognition)(execution),
            label={[font=\small\scshape, yshift=-2pt]above:\sys Security Kernel}] {};
    \end{scope}

    \node[left=1.2cm of identity, font=\small, align=center,
          text=red!70!black] (input) {Input\\World\\[-2pt]\scriptsize(untrusted)};
    \node[right=1.2cm of execution, font=\small, align=center,
          text=green!50!black] (output) {Output\\World\\[-2pt]\scriptsize(auditable)};

    \draw[arrow, red!60!black]  (input)  -- (identity.west);
    \draw[arrow, red!60!black]  (identity.east) -- (perception.west);
    \draw[arrow, green!50!black] (execution.east) -- (output);
  \end{tikzpicture}
  \caption{%
    \sys architecture.  The security kernel mediates all interactions
    between the untrusted input world and the auditable output world.
    Four security pillars operate around the protected agent core,
    each enforcing independent invariants.%
  }
  \label{fig:architecture}
\end{figure*}

\sys is our instantiation of this contract.  It is realized as a
\emph{mandatory enforcement boundary} wrapped around the agent core: every
interaction---inbound content, outbound actions, agent-to-agent
communication---must traverse the security kernel, with each pillar
enforcing independent invariants.
Figure~\ref{fig:architecture} illustrates the overall architecture.

The architecture is governed by two standalone trusted services that
operate outside the agent's control:

\begin{itemize}
  \item The \textbf{Global Agent Registry (GAR)}---a remote,
    CA-grade service responsible for developer enrollment, Agent Identity
    Card (AIC) issuance via Ed25519 signatures, policy governance, and
    revocation.
  \item The \textbf{Agent Kernel}---a local trust anchor that holds
    the agent's private key (never exported), mediates mutual
    attestation, manages OIDC secrets, and hosts pluggable security
    modules for each pillar.
\end{itemize}

The agent application connects to \sys through exactly three narrow adapter
interfaces (Principle~5)---bounding the agent-visible surfaces of model
inference, tool execution, and memory I/O: the LLM adapter, the tool
adapter, and the storage adapter.  This confined integration boundary
ensures that the security kernel can be independently audited and
replaced without modifying agent logic.

\subsection{Pillar 1: Identity as a Kernel-Managed Resource}
\label{sec:arch:identity}

\begin{insight}
Agent identity must be a kernel-managed resource---held, protected,
and mediated by the security kernel, never exposed to agent application
code.  Just as an OS kernel manages process credentials on behalf of
user-space programs, the agent security kernel manages cryptographic
identity on behalf of agents.
\end{insight}

In existing agent frameworks, identity is a self-declared string---an
agent name or an API key---that lives in the agent's own configuration
and is fully accessible to application code.  This conflation of
identity \emph{holder} and identity \emph{subject} is the root cause
of agent impersonation, unauthorized delegation, and supply-chain
attacks: the agent that uses a credential is the same entity that
stores it, so compromising the agent automatically compromises the
identity.

\begin{figure*}[t]
  \centering
  \input{figures/identity-kernel}
  \caption{%
    Identity as a kernel-managed resource.  The Agent Kernel holds the
    private key and mediates all identity operations on behalf of the
    agent.  Identity-derived trust signals flow into each of the other
    three security pillars through the kernel's control plane, making
    identity the shared trust substrate for the entire security
    architecture.%
  }
  \label{fig:identity-kernel}
\end{figure*}
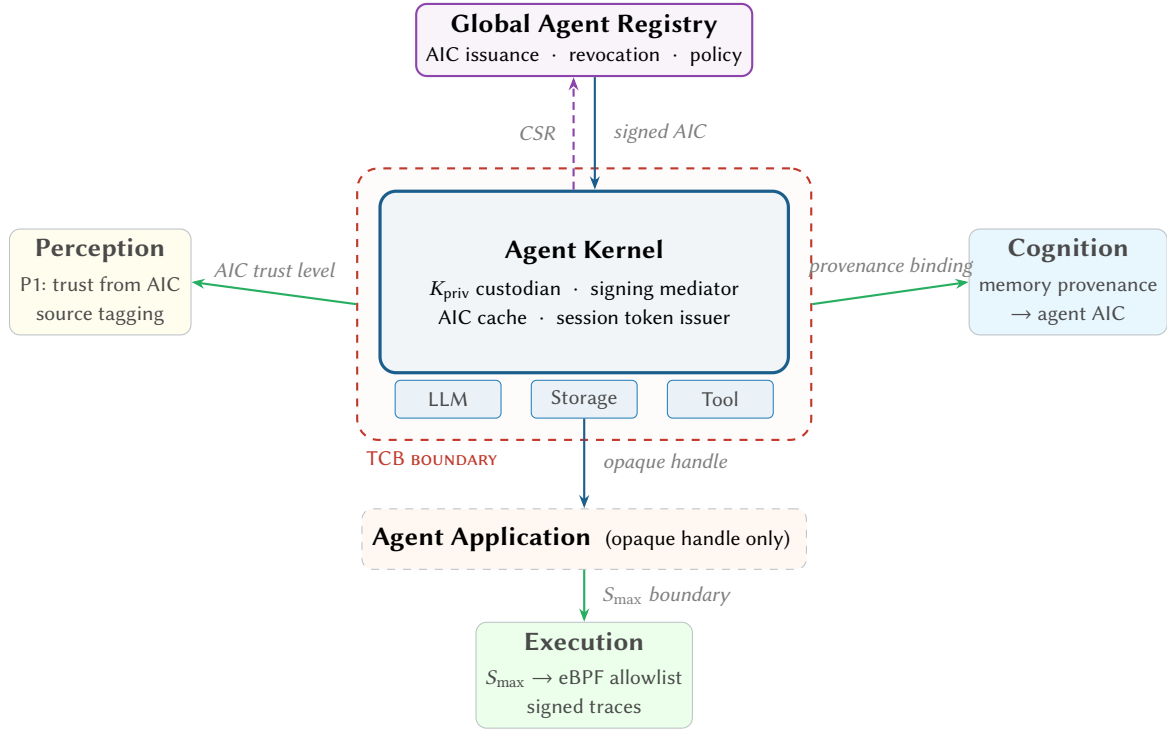

\sys breaks this conflation by treating identity as a
\emph{kernel-managed resource}, analogous to how an OS kernel manages
process UIDs and capabilities.  The agent's private key $\Kpriv$
resides exclusively within the Agent Kernel---a local trust anchor
that acts as an HSM-like custodian.  Agent application code never
touches raw key material; it interacts only with opaque signing
handles.  All cryptographic operations---AIC presentation, challenge
signing, session token issuance---are \emph{mediated by the kernel},
not performed by the agent itself. Figure~\ref{fig:identity-kernel} illustrates this
kernel-mediated identity architecture.

\paragraph{Provisioning: what the kernel binds.}
At provisioning time, the Agent Kernel generates a fresh Ed25519
keypair, constructs an Agent Identity Card (AIC) that
cryptographically binds four dimensions---developer, code artifact,
operator, and deployment context---into a single verifiable
credential~\cite{w3c-vc-2.0}, and submits it to the Global Agent
Registry (GAR) for signing.  The signed AIC is cached locally; the
private key never leaves the kernel boundary.  The four-dimensional
binding model builds on our earlier work on cryptographic agent
identity~\cite{blocka2a}; here we focus on how the kernel \emph{uses}
this identity to drive OS-level enforcement.

\paragraph{TEE-backed key custody.}
Identity-key storage follows a \emph{tiered} custody path---from
file- and OS-keychain--backed handles up to an optional trusted
execution environment (TEE) tier (e.g., Intel SGX, ARM TrustZone, or
Nitro-class enclaves)~\cite{sgxexplained,ngabonziza2016trustzone}.
When the TEE tier is used, the identity private key is provisioned and
used \emph{inside} the enclave so raw material never crosses the
hardware isolation boundary; the Agent Kernel still mediates signing
through non-exportable primitives, preserving the same
kernel-controlled interface to agent code while strengthening
extraction resistance against a compromised host OS.

\paragraph{Identity-driven enforcement.}
In a traditional OS, a process's UID determines which files it can
open, which signals it can send, and which system calls it can
invoke.  In \sys, the AIC's \emph{capability boundary} $\Smax$
plays an analogous role: it is the structural input to every
enforcement decision across the security kernel.
At the execution pillar, Layer~E1 cross-references each tool
invocation against the intersection of the agent's $\Smax$ and
the tool's declared permission manifest; Layer~E3 translates this
intersection into an eBPF allowlist installed before the tool's
process tree is spawned.  When multiple skills compose within a
single agent, the kernel computes the \emph{intersection} of all
skill permission envelopes with the agent's $\Smax$
(Principle~2)---ensuring that composition can only narrow, never
widen, the effective capability set.

\paragraph{Cross-pillar identity integration.}
Unlike protocol-level identity systems that operate in isolation,
\sys's kernel-managed identity serves as a \emph{shared trust
substrate} consumed by all four security pillars:

\begin{itemize}
  \item \textbf{Perception.}  Layer~P1 (Source Tagger) assigns
    baseline trust levels partly based on whether the content source
    carries a valid AIC: messages from AIC-verified agents receive
    higher initial trust than content from unverified origins,
    reducing perception overhead for trusted peers.
  \item \textbf{Cognition.}  Every memory item's provenance chain
    traces back to the originating agent's AIC, enabling the memory
    gateway to enforce label-based access control---an agent
    can retrieve only items whose provenance labels are compatible
    with its own identity and clearance level.
  \item \textbf{Execution.}  As described above, the AIC's capability
    boundary is the structural ceiling for all tool-call permissions
    and eBPF allowlists.  Additionally, execution traces are signed
    by the kernel using the agent's $\Kpriv$, binding every
    recorded action to a non-repudiable identity.
\end{itemize}

This cross-pillar integration is the key architectural difference
between \sys's identity pillar and standalone agent identity
protocols: identity is not merely an authentication credential
exchanged at session start, but a \emph{live enforcement input}
that constrains every security decision throughout the agent's
lifecycle.

\paragraph{Kernel-mediated trust establishment.}
When two agents need to interact, neither agent touches the
cryptographic handshake directly.  Instead, the Agent Kernel
orchestrates the entire trust negotiation: it generates fresh
nonces, verifies both parties' AICs against the GAR trust store,
signs challenges using the kernel-held $\Kpriv$, and issues a
scoped session token whose capability boundary is the intersection
$\Ssession \subseteq \Smax^A \cap \Smax^B$.  The token is
context-bound and lifecycle-ephemeral---it expires when either
agent terminates, preventing credential leakage.  Because all
signing is kernel-mediated, a compromised agent cannot forge
attestations or escalate its own token scope, just as a compromised
user-space process cannot forge kernel-issued capabilities.

\paragraph{Delegation as a kernel operation.}
In multi-agent orchestration, spawning a sub-agent with delegated
authority is a \emph{kernel system call}, not an application-level
action.  The kernel verifies that the child's requested capabilities
are a strict subset of the parent's $\Smax$, enforces a configurable
depth ceiling, and issues a child AIC whose validity is capped to the
parent's remaining lifetime.  Revoking a parent AIC atomically
invalidates the entire delegation subtree---cascading revocation is
enforced structurally by the kernel, not by policy.

\subsection{Pillar 2: Perception}
\label{sec:arch:perception}

\begin{insight}
The best defense intercepts threats before the LLM ever sees them.
Defense-in-depth should apply across layers of increasing cost and
sophistication.
\end{insight}

LLM-powered agents ingest content from inherently untrusted sources.
The perception pillar establishes a \emph{graduated defense pipeline}
that processes all input before it enters the LLM's context window.
Unlike single-point defenses, \sys's perception is designed as
\emph{layered depth}: each layer catches what the previous one misses,
with cost and sophistication increasing at each stage.

\paragraph{Layer P1---Source Tagger.}
Every piece of inbound content is tagged with a \emph{trust level}
based on its origin: \texttt{SYSTEM} (fully trusted),
\texttt{USER} (high trust), \texttt{TOOL} (conditional), \texttt{EXTERNAL}
(low---boundary markers added), or \texttt{UNTRUSTED} (flagged by
downstream filters).  These tags generate \emph{ProvenanceTag} metadata
(origin, timestamp, trust level, detection warnings) that persist through
the entire pipeline and are consumed by downstream cognition and
execution pillars.

\paragraph{Layer P2---Rule-Based Filter.}
A high-speed pattern-matching layer intercepts known injection
templates: instruction overrides (``ignore previous instructions''),
role impersonation, encoding bypasses (Base64, Unicode obfuscation), and
token-stuffing payloads.  Rules are sub-millisecond, deterministic, and
auditable, handling the majority of known attack patterns at near-zero
cost.

\paragraph{Layer P3---Semantic Firewall.}
For content that passes rule-based filters but may contain semantically
adversarial intent, a lightweight LLM classifier evaluates whether the
content attempts to manipulate agent behavior, override system
instructions, or present intent inconsistent with its declared type.
The classifier operates as a binary decision gate with context-aware
thresholds---stricter for low-trust sources identified by P1.

\paragraph{Layer P4---Jailbreak and Multi-Turn Detector.}
Sophisticated attacks unfold across multiple conversation
turns~\cite{shen2024jailbreak}.  P4 analyzes conversation history to
detect gradual role redefinition, progressive safety degradation,
statistical deviations from normal flow, and known jailbreak templates.

\paragraph{Integration with identity.}
Provenance tags inherit trust levels from verified agent identities:
content from agents with valid AICs receives higher baseline trust than
content from unverified sources.  This creates a virtuous cycle where
strong identity reduces perception overhead for trusted agents.

\subsection{Pillar 3: Cognition}
\label{sec:arch:cognition}

\begin{insight}
Memory is not just the agent's persistent context---it is a security
boundary.  Information flow control theory, applied at entry granularity,
can preserve usability while preventing contamination.
\end{insight}

As agents gain long-term memory and multi-agent collaboration
capabilities, memory becomes a critical security surface.  A single
piece of untrusted content persisted in memory can \emph{persistently
influence behavior across sessions}---far more dangerous than transient
prompt injection.

\sys's cognition pillar treats memory as a \emph{managed, labeled,
auditable data plane} with information-flow guarantees, structured
across four layers: Agent Host, Memory Gateway, Memory Control
Plane, and Storage.

\paragraph{Lattice-based entry-level taint propagation.}
The core algorithmic innovation is \emph{item-level label propagation
on a product lattice}, rather than coarse-grained ``entire session
inherits the worst label'' approaches.

Consider a session containing: a user preference ($\mathit{trusted}
\times \mathit{public}$), a private email ($\mathit{trusted} \times
\mathit{secret}$), and a web excerpt ($\mathit{untrusted} \times
\mathit{public}$).  Naively taking the join produces
$\mathit{untrusted} \times \mathit{secret}$ for the \emph{entire}
session---rendering most memory unusable.

\sys instead operates at the individual memory item level.  The
process follows five steps:
\emph{(1)}~segment the conversation into source segments with known
provenance;
\emph{(2)}~extract candidate memory items;
\emph{(3)}~for each candidate, determine which source segments
actually support it;
\emph{(4)}~search the label lattice for the minimum security label
that still supports the item's validity (utility loss $\leq \lambda$);
\emph{(5)}~output a set of incomparable minimal labels (a minimal
anti-chain), not a single worst-case collapse.

This approach---rooted in classical lattice models of information
flow~\cite{denning1976lattice,bell1973secure,biba1977integrity}---preserves
memory usability while maintaining provenance precision.

\paragraph{Cross-agent memory sharing.}
\sys adopts a ``default private, explicit sharing'' principle.  Memory
is private by default; sharing requires explicit authorization
(\texttt{shared\_with} lists, task-scoped workspaces, or human-approved
export).  Memory with high-sensitivity labels is \emph{never
automatically shared} across agent boundaries.  Even shared memory
maintains independent access audit trails per agent.

\paragraph{Anti-injection for memory.}
Historical context retrieved from memory is explicitly marked as
\emph{non-executable data}---it cannot be interpreted as instructions.
Combined with provenance chains that trace every memory entry to its
origin, this prevents the ``memory-as-instruction'' attack pattern.

\subsection{Pillar 4: Execution}
\label{sec:arch:execution}

\begin{insight}
From intent to syscall, every layer must enforce least privilege.
Claimed actions must be verified against actual execution traces.
\end{insight}

The execution pillar bridges the gap between \emph{what the model says}
and \emph{what the system actually does}, through four layers that
collaborate to ensure integrity.

\paragraph{Layer E1---Rule/Policy Evaluator.}
At the tool invocation boundary, fast deterministic checks enforce
deny-lists, capability alignment (cross-referencing the tool's manifest
with the agent's AIC boundary), path/command/domain constraints, and
deny-first conflict resolution.  Cost: sub-millisecond, no LLM call.

\paragraph{Layer E2---LLM Validator.}
For high-ambiguity or high-impact decisions where rules alone are
insufficient, a lightweight LLM evaluates intent alignment (``Is this
action consistent with the user's stated goal?'') and may trigger
human-in-the-loop confirmation.  Invoked only when E1 escalates.

\paragraph{Layer E3---eBPF/Kernel Hook Manager.}
This is \sys's most distinctive execution innovation.  Rather than
trusting tool implementations to respect declared constraints, a
\emph{privileged guardian daemon}:
\emph{(1)}~pre-computes an allowlist before each tool execution (which
file paths, network endpoints, and DNS queries are permitted);
\emph{(2)}~installs the allowlist in kernel-level eBPF probes or system
hooks;
\emph{(3)}~monitors the entire process tree spawned by the tool---child
processes inherit the same constraints; and
\emph{(4)}~pauses violating processes immediately, then applies
remediation.
Even if a tool's implementation is malicious, it \emph{cannot} access
resources outside the pre-computed allowlist.

\paragraph{Layer E4---Plan--Trace Aligner.}
Post-execution verification aligns what was planned with what actually
happened.  Every tool call's parameters, timing, and result are recorded
as an immutable trace.  The aligner compares the trace against the
agent's stated plan, surfacing missing steps, extra steps, reordered
steps, and---critically---\emph{hallucinated actions} where the LLM
claims to have performed operations that the trace shows never occurred.

\paragraph{Skill permission manifests \& workflow orchestration.}
Tools and skills declare their required system surfaces (file system,
network, execution, messaging) via structured manifests.  When multiple
skills compose into a workflow, \sys computes the \emph{intersection}, not the union,
of their permission sets (Principle~2).  The
declaration layer determines ``how far a skill can reach at most'';
the runtime layer (\mbox{E1--E3}) determines ``whether this specific
invocation can actually happen.''
This structural permission convergence provides a secure substrate for
\emph{tool and workflow orchestration}: orchestrators can freely chain
tools and delegate sub-tasks, knowing that the kernel will automatically
bound the workflow's authority to the safest common denominator.

\paragraph{Agent scheduling \& concurrency isolation.}
As agents scale to handle concurrent workflows, \sys provides the necessary
isolation primitives for \emph{agent scheduling and concurrency}.
Consistent with the identity pillar (\S\ref{sec:arch:identity}), each concurrent
execution context is cryptographically bound: local concurrent threads receive
distinct, lifecycle-ephemeral execution tokens, while delegated sub-agents
operate under their own child AICs.
At the execution layer, Layer~E3 enforces concurrency isolation by mapping
each execution token or child AIC to a dedicated BPF namespace and process tree.
This ensures that concurrent agent tasks---even those running the same
skill---cannot interfere with each other's memory, file descriptors, or
network sockets, enabling safe, high-throughput agent scheduling.

\paragraph{Deterministic-first philosophy.}
The main execution path uses deterministic allowlists and kernel
enforcement; LLM adjudication (E2) is an optional enhancement for edge
cases.  This ensures that security does not degrade when the LLM makes
errors or is manipulated.

%% file: figures/identity-kernel.tex
\begin{tikzpicture}[
  kernel/.style={
    draw=dkblue, very thick, rounded corners=6pt,
    minimum width=5.4cm, minimum height=2.4cm,
    fill=dkblue!6, align=center, font=\sffamily\small
  },
  pillar/.style={
    draw=dkborder, rounded corners=4pt,
    minimum width=2.4cm, minimum height=1.4cm,
    align=center, font=\sffamily\small, fill=white, text=dkdarkgray
  },
  remote/.style={
    draw=dkpurple, thick, rounded corners=4pt,
    minimum width=3.4cm, minimum height=0.9cm,
    fill=dkpurple!6, align=center, font=\sffamily\small
  },
  agent/.style={
    draw=dkborder, dashed, rounded corners=4pt,
    minimum width=2.8cm, minimum height=0.8cm,
    fill=dkorange!6, align=center, font=\sffamily\small
  },
  port/.style={
    draw=dkaccent, rounded corners=2pt,
    minimum width=1.4cm, minimum height=0.5cm,
    fill=dkaccent!10, align=center,
    font=\sffamily\scriptsize, text=dkdarkgray, inner sep=2pt
  },
  arr/.style={-{Stealth[length=5pt]}, thick, color=dkblue},
  darr/.style={-{Stealth[length=5pt]}, thick, color=dkpurple,
               densely dashed},
  tarr/.style={-{Stealth[length=5pt]}, thick, color=dkgreen},
  lbl/.style={font=\sffamily\scriptsize\itshape, text=gray},
]

\node[kernel] (kern) at (0,0) {
  \textbf{Agent Kernel}\\[2pt]
  {\scriptsize $\Kpriv$ custodian\; $\cdot$\; signing mediator}\\[-1pt]
  {\scriptsize AIC cache\; $\cdot$\; session token issuer}
};

\node[port] (llmport)  at (-1.8, -1.55) {LLM};
\node[port] (storport) at ( 0.0, -1.55) {Storage};
\node[port] (toolport) at ( 1.8, -1.55) {Tool};

\node[draw=dkred, dashed, thick, rounded corners=8pt,
      inner sep=8pt,
      fit=(kern)(llmport)(storport)(toolport),
      fill=none] (tcb) {};
\begin{scope}[on background layer]
  \fill[dkred!2, rounded corners=8pt] (tcb.south west) rectangle (tcb.north east);
\end{scope}
\node[font=\sffamily\scriptsize\scshape, text=dkred,
      anchor=north west] at (tcb.south west) {TCB boundary};

\node[remote] (gar) at (0, 3.2) {
  \textbf{Global Agent Registry}\\[-1pt]
  {\scriptsize AIC issuance\; $\cdot$\; revocation\; $\cdot$\; policy}
};

\node[agent] (app) at (0, -3.4) {
  \textbf{Agent Application}\;
  {\scriptsize (opaque handle only)}
};

\node[pillar, fill=yellow!8] (percep) at (-6.4, 0) {
  \textbf{Perception}\\[1pt]
  {\scriptsize P1: trust from AIC}\\[-1pt]
  {\scriptsize source tagging}
};

\node[pillar, fill=cyan!8] (cognit) at (6.4, 0) {
  \textbf{Cognition}\\[1pt]
  {\scriptsize memory provenance}\\[-1pt]
  {\scriptsize $\to$ agent AIC}
};

\node[pillar, fill=green!8] (execut) at (0, -5.2) {
  \textbf{Execution}\\[1pt]
  {\scriptsize $\Smax \to$ eBPF allowlist}\\[-1pt]
  {\scriptsize signed traces}
};

\draw[arr]  ([xshift=4pt]gar.south) -- ([xshift=4pt]kern.north)
  node[lbl, midway, right=3pt] {signed AIC};
\draw[darr] ([xshift=-4pt]kern.north) -- ([xshift=-4pt]gar.south)
  node[lbl, midway, left=3pt] {CSR};

\draw[arr] (storport.south) -- (app.north)
  node[lbl, midway, right=3pt] {opaque handle};

\draw[tarr] (tcb.west) -- (percep.east)
  node[lbl, midway, above=2pt] {AIC trust level};

\draw[tarr] (tcb.east) -- (cognit.west)
  node[lbl, midway, above=2pt] {provenance binding};

\draw[tarr] (app.south) -- (execut.north)
  node[lbl, midway, right=3pt] {$\Smax$ boundary};

\end{tikzpicture}

%% file: sections/analysis.tex
\section{Security Analysis}
\label{sec:analysis}

A security architecture is only as strong as the argument that supports it.
This section states the threat model and trust assumptions under which
\sys's guarantees hold (\S\ref{sec:analysis:threat}), derives precise
security invariants for each pillar and argues their soundness
(\S\ref{sec:analysis:invariants}), examines how the four pillars compose
into end-to-end guarantees (\S\ref{sec:analysis:composition}), analyzes
three architectural choices that fundamentally distinguish \sys's
security model from prior approaches (\S\ref{sec:analysis:distinctions}),
and closes with an honest accounting of residual risks
(\S\ref{sec:analysis:residual}).

\subsection{Threat Model and Trust Assumptions}
\label{sec:analysis:threat}

\sys targets a strong adversary model in which the attacker may control
inputs, supply compromised tools, and exploit misconfigurations, but
cannot break standard cryptographic assumptions or bypass the mandatory
mediation boundary.

\parab{Attacker capabilities}
We assume an adversary with the following capabilities:
\begin{enumerate}[leftmargin=2em]
  \item \textbf{Input manipulation.}  The attacker can craft arbitrary
    inputs to the agent---prompt injection in tool outputs, web pages,
    files, and inter-agent messages; jailbreak attempts; encoding bypasses;
    and multi-turn conversational attacks.  These attacks target the
    perception and cognition pillars.
  \item \textbf{Tool compromise.}  The attacker can supply a malicious tool
    that passes semantic-level parameter inspection but attempts
    unauthorized operations at runtime---spawning child processes,
    accessing files outside declared scope, or opening network connections
    to exfiltration endpoints.  This targets the execution pillar.
  \item \textbf{Memory poisoning.}  The attacker can inject content into
    agent memory through normal interaction channels, aiming to influence
    retrieval and decision-making in later sessions
    (\S\ref{sec:bg:threats}).  This targets the cognition pillar.
  \item \textbf{Impersonation and replay.}  The attacker can attempt to
    masquerade as a legitimate agent or operator using replayed
    credentials, self-declared identity strings, or stolen API keys.
  \item \textbf{Supply-chain injection.}  The attacker can publish
    malicious skills or plugins that declare deceptively narrow
    permissions while containing hidden malicious functionality.
  \item \textbf{Operator misconfiguration.}  A benign but mistaken
    operator may set overly permissive policies.  \sys's structural
    guarantees are designed to survive this case
    (\S\ref{sec:analysis:structural}).
\end{enumerate}

\parab{Attacker limitations and trust assumptions}
The following capabilities are assumed \emph{outside} the attacker's reach;
violating any of them voids the corresponding guarantees:
\begin{enumerate}[leftmargin=2em]
  \item The attacker \textbf{cannot forge Ed25519 signatures} without
    access to the corresponding private key, nor break standard
    cryptographic assumptions (hash collision resistance, etc.).
  \item The attacker \textbf{cannot extract private keys from the Agent
    Kernel.}  The kernel holds $\Kpriv$ and mediates all signing
    operations; agent application code never touches raw key material
    (\S\ref{sec:arch:identity}).
  \item The attacker \textbf{cannot bypass the three adapter interfaces}
    (LLM, tool, storage).  All agent-visible paths to model inference,
    tool execution, and durable memory must traverse these adapters
    (Principle~5).  If the agent retains ambient credentials or direct
    filesystem access outside the adapters, mediation is incomplete.
  \item The attacker \textbf{cannot modify eBPF probes} without OS
    kernel-level privilege.  The host OS kernel is assumed to correctly
    enforce the allowlists installed by Layer~E3
    (\S\ref{sec:arch:execution}).
  \item The \textbf{GAR signing key is assumed uncompromised.}  GAR is the
    root of trust for agent identity; its compromise would enable
    arbitrary AIC forgery and is therefore catastrophic for the identity
    pillar.
\end{enumerate}

\parab{Explicitly out of scope}
Physical attacks, side-channel attacks (timing, power analysis), and compromise of the
host OS kernel itself are outside the threat model.  Attacks that exploit
LLM reasoning errors without violating a stated invariant (e.g., the agent
making a poor but policy-compliant decision) are reliability issues rather
than security violations in our model.

\parab{Trusted computing base}
The TCB consists of the Global Agent Registry (GAR), the Agent Kernel, the
three adapter interfaces, the OS kernel's eBPF subsystem, and the
cryptographic primitives (Ed25519, hash functions).  Each component is
described in \S\ref{sec:architecture}; their correctness is the foundation
on which the following invariants rest.

\subsection{Security Invariants}
\label{sec:analysis:invariants}

For each pillar we state precise security invariants, identify the
mechanism that enforces them, and give a brief soundness argument.
Table~\ref{tab:security-props} provides a consolidated summary.

\begin{table}[t]
  \centering
  \caption{Security invariants of the \sys architecture.}
  \label{tab:security-props}
  \small
  \begin{tabularx}{\linewidth}{@{}>{\raggedright\arraybackslash}p{1.5cm} >{\raggedright\arraybackslash\hsize=0.76\hsize}X >{\raggedright\arraybackslash\hsize=1.24\hsize}X@{}}
    \toprule
    \textbf{Pillar} & \textbf{Invariant} & \textbf{Soundness Argument} \\
    \midrule
    \multirow{4}{*}{Identity}
      & I1. AIC unforgeability &
        Ed25519 signatures verified against GAR root at every presentation;
        $\Kpriv$ never leaves kernel boundary (\S\ref{sec:arch:identity}). \\[3pt]
      & I2. Delegation monotonicity: $\Smax^{\text{child}} \subseteq \Smax^{\text{parent}}$ &
        Kernel verifies subset relation before issuing child AIC; child
        validity is capped to parent's remaining lifetime. \\[3pt]
      & I3. Action non-repudiation &
        All execution traces kernel-signed by $\Kpriv$; signing is
        kernel-mediated, never performed by agent code. \\[3pt]
      & I4. Session anti-replay &
        Fresh nonces with TTL per session; context-bound tokens;
        lifecycle ephemerality (\S\ref{sec:arch:identity}). \\
    \cmidrule{1-3}
    Perception
      & P1. Provenance completeness: all content
        entering LLM context carries origin tags &
        All input paths traverse P1 (Source Tagger) as mandatory first
        stage; tags persist through P2--P4. \\[3pt]
      & P2. Layered independence: bypassing one
        filter does not help bypass others &
        P1--P4 operate on orthogonal features (source identity, syntactic
        patterns, semantic intent, multi-turn history). \\
    \cmidrule{1-3}
    Cognition
      & C1. Taint soundness: derived items inherit
        dominating labels from source segments &
        Lattice search identifies all supporting segments and computes
        minimal dominating label (\S\ref{sec:arch:cognition}). \\[3pt]
      & C2. Default-private sharing &
        Memory gateway enforces explicit authorization on every
        cross-agent access; no implicit sharing. \\
    \cmidrule{1-3}
    Execution
      & E1. Intersection least-privilege:
        $S_{\text{eff}} = \bigcap_i S_i$ &
        Kernel computes intersection at each composition point
        (Principle~2); deny-by-default resolution. \\[3pt]
      & E2. eBPF non-bypassability: no process
        exceeds dynamically installed allowlist &
        eBPF probes installed before process spawn; child processes
        inherit constraints (\S\ref{sec:arch:execution}, E3). \\[3pt]
      & E3. Plan--trace alignment: deviations
        between declared plan and execution are detected &
        E4 records immutable traces and compares against stated plan;
        hallucinated actions surface as mismatches. \\
    \bottomrule
  \end{tabularx}
\end{table}

\parab{Identity invariants}
The identity pillar provides the cryptographic substrate on which all
other pillars depend.  I1 (AIC unforgeability) follows from Ed25519
security and the fact that GAR is the sole authority for AIC issuance;
an agent cannot self-certify an identity.  I2 (delegation monotonicity)
is the structural guarantee that distinguishes \sys's capability model:
the kernel enforces $\Smax^{\text{child}} \subseteq
\Smax^{\text{parent}}$ by construction, and no policy override can widen
the child's boundary.  I3 and I4 are standard cryptographic properties
(non-repudiation and freshness) realized through kernel-mediated signing
and single-use nonces respectively.

\parab{Perception invariants}
P1 is a \emph{mediation completeness} property: every unit of content
that reaches the LLM context must carry a provenance tag.  It holds
because all input paths (user messages, tool outputs, web content,
inter-agent messages) are routed through the graduated perception
pipeline, where P1 (Source Tagger) is the mandatory first stage.  P2 is
a \emph{defense-in-depth} property: the four layers P1--P4 operate on
orthogonal features, so an attacker who evades one layer (e.g., crafting a
payload that passes syntactic filters) still faces independent barriers
at the semantic and temporal layers.

\parab{Cognition invariants}
C1 is an \emph{information-flow soundness} property.  The lattice-based
search at item granularity (\S\ref{sec:arch:cognition}) ensures that the
taint label assigned to each memory item dominates the labels of all
source segments from which it was derived.  Critically, this guarantee
depends on the accuracy of source segmentation (step~1 of the algorithm);
adversarially crafted content designed to blur segment boundaries can
degrade label fidelity---a limitation we address in
\S\ref{sec:analysis:residual}.  C2 enforces that cross-agent memory
sharing is explicit and auditable, preventing the ``contaminated memory
propagates silently'' failure mode identified in
\S\ref{sec:bg:threats}.

\parab{Execution invariants}
E1 is the most architecturally consequential invariant: it guarantees
that policy composition can only \emph{narrow} permissions.  Unlike
systems where multiple policy sources are combined via union or priority
ordering, \sys's intersection semantics ensure that no single permissive
policy can override a restrictive one.  E2 bridges the gap from semantic
permission to syscall enforcement: the dynamically computed allowlist is
installed as eBPF probes before tool execution, and the entire process
tree inherits the constraints.  E3 provides \emph{detection} rather than
prevention: it does not block plan--trace mismatches but guarantees they
are surfaced for audit, enabling retrospective accountability even when
preventive controls fail.

\subsection{Composition and Cross-Pillar Guarantees}
\label{sec:analysis:composition}

The four pillars are not isolated; they share provenance metadata that
flows through the system along a defined chain: Identity establishes
\emph{who} produced content; Perception assigns initial \emph{trust
levels}; Cognition propagates \emph{taint labels} through memory
derivations; Execution consumes these labels to build \emph{allowlists}
and constrain tool actions.  This flow creates both composition benefits
and coupling risks, which we analyze in turn.

\parab{Layered defense}
The key composition argument is that an attacker must defeat
\emph{multiple independent mechanisms} to achieve end-to-end compromise.
Consider the motivating DevOps scenario from
\S\ref{sec:intro:scenario}: to successfully push malicious code, an
attacker must (i)~evade identity checks to impersonate a legitimate
delegator, (ii)~bypass perception filters to deliver the injection
payload, (iii)~overcome taint propagation in cognition to have the
poisoned summary treated as trusted context, and (iv)~execute tool
actions that pass both semantic alignment (E1--E2) and syscall
enforcement (E3).  Each pillar independently reduces the residual attack
surface: even if perception misses an injection, cognition's taint labels
still flag the resulting memory entry; even if cognition propagates
contaminated labels, execution's eBPF allowlist remains constrained by
the agent's $\Smax$ and the tool's declared manifest.

\parab{Coupling and independence}
We explicitly do \emph{not} claim that the four pillars are fully
independent---they share provenance metadata, so a failure in an upstream
pillar can affect downstream decisions.  If identity verification is
compromised (I1 violated), perception assigns incorrect trust levels; if
source segmentation fails, cognition computes incorrect taint labels; if
taint labels are wrong, execution may construct overly permissive
allowlists.  What the architecture guarantees is not independence of
\emph{outcomes} but independence of \emph{enforcement mechanisms}: each
pillar independently re-validates upstream claims against its own
invariants rather than blindly trusting upstream metadata.  Perception
cross-checks identity claims; cognition cross-checks provenance tags
against source segments; execution cross-checks taint labels against
declared manifests.  This \emph{re-validation chain} means that an
upstream failure must survive downstream scrutiny to become an end-to-end
compromise---a higher bar than a single-point failure.

\parab{Why four pillars, not one}
A natural question is whether a single unified policy engine could
achieve the same guarantees with less complexity.  The answer lies in the
\emph{heterogeneity of the attack surface}: identity is a cryptographic
problem; perception is a classification problem; cognition is an
information-flow problem; execution is a mediation problem.  Each demands
different mechanisms, failure models, and verification techniques.
Unifying them into a single layer would force a least-common-denominator
approach that weakens guarantees at every boundary.  The four-pillar
decomposition is not an aesthetic choice; it follows from the structure
of the threat surface itself.

\subsection{Architectural Distinctions}
\label{sec:analysis:distinctions}

Three architectural choices in \sys yield security properties that are
unattainable under conventional designs.  We analyze each as a security
argument rather than a design rationale.

\subsubsection{Structural vs.\ Policy-Based Least-Privilege}
\label{sec:analysis:structural}

Most agent frameworks enforce least-privilege through policy engines
(OPA, Cedar, XACML).  Policy-based approaches share a structural
weakness: they \emph{fail open} on misconfiguration.  An overly
permissive policy silently grants excessive access, and the system has no
independent means to detect the error because the policy \emph{is} the
authority.

\sys inverts this failure mode through cryptographically enforced
monotonic attenuation.  At each stage---developer declaration, operator
provisioning, GAR issuance, runtime enforcement---the effective
capability set is the \emph{intersection} of the current stage's grant
with the cryptographically signed boundary from the preceding stage:
\[
  \Smax^{\text{child}} \;\subseteq\; \Smax^{\text{parent}}
  \qquad\text{(enforced by signature verification, not policy evaluation)}
\]
No policy change at any single stage can grant capabilities beyond the
preceding stage's signed envelope.  We state this as a structural
guarantee:

\begin{property}[Monotonic Capability Attenuation]
\label{prop:monotonic}
For any delegation chain $A_0 \to A_1 \to \cdots \to A_n$, the
effective capability set satisfies
$\Smax^{A_n} \subseteq \Smax^{A_{n-1}} \subseteq \cdots \subseteq \Smax^{A_0}$.
Policy can \emph{deny} within this envelope but never \emph{grant} beyond it.
\end{property}

The guarantee follows from two architectural invariants: (i)~the kernel
verifies the subset relation $\Smax^{\text{child}} \subseteq
\Smax^{\text{parent}}$ against the parent's signed AIC before issuing a
child AIC, and (ii)~runtime enforcement (E1) further constrains effective
permissions to the intersection of all applicable policies.  Even a
misconfigured operator policy cannot override invariant~(i) because it is
enforced at the cryptographic layer, not the policy layer.

The practical implication is significant: in a multi-stakeholder
deployment, no single stakeholder can unilaterally escalate privileges.
This is the property that enables safe cross-organizational agent
delegation---a capability that policy-only approaches cannot provide
without a trusted enforcement layer external to the policy engine.

\subsubsection{Item-Level vs.\ Session-Level Taint}
\label{sec:analysis:taint}

Classical information-flow control~\cite{denning1976lattice} operates at
the granularity of containers (files, processes).  When applied to agent
memory, the natural container is the \emph{session}: all items derived
from a session inherit its worst-case label.  This approach is
\emph{sound} (no false negatives) but \emph{unusable} (excessive false
positives)---a single untrusted web excerpt taints the entire session's
memory, rendering most items inaccessible.

\sys resolves this tension by operating at \emph{item granularity} on a
product lattice (integrity $\times$ confidentiality).  The algorithm
(\S\ref{sec:arch:cognition}) identifies the \emph{minimal} label that
preserves each item's validity, producing a set of incomparable minimal
labels (an anti-chain) rather than a single worst-case join.  The
security question is whether item-level granularity preserves soundness.

The soundness of item-level taint rests on the accuracy of source
segmentation (identifying which source segments support each candidate
memory item).  When segmentation is correct, the item-level label
dominates all contributing sources by construction (invariant~C1).  When
segmentation is imperfect---as can happen with adversarially crafted
content designed to blur provenance boundaries---the label may
under-approximate the true taint.  This is a \emph{false-negative risk}
that session-level taint avoids entirely.

The architectural trade-off is therefore: session-level taint is sound
but yields a degenerate lattice (one label per session), making
information-flow control impractical for interactive agents; item-level
taint trades a bounded amount of soundness (under adversarial
segmentation) for dramatically improved usability.  \sys accepts this
trade-off explicitly, with the understanding that the execution pillar
(E2--E3) provides an independent backstop: even if cognition
under-labels a memory item, the eBPF allowlist derived from the agent's
$\Smax$ and the tool's manifest remains in force.

\subsubsection{The Semantic--Syscall Bridge}
\label{sec:analysis:bridge}

The execution pillar addresses a fundamental architectural gap that
neither semantic nor syscall enforcement can close alone:
\begin{itemize}
  \item \textbf{Semantic-only enforcement is bypassable.}  A malicious
    tool that passes parameter inspection can still spawn child processes
    or access resources outside its declared scope---the tool's
    implementation is a black box at the semantic level.
  \item \textbf{Syscall-only enforcement is semantically blind.}  It
    cannot distinguish a file write driven by legitimate reasoning from
    one driven by a poisoned memory entry---both are identical at the
    syscall boundary.
\end{itemize}

\sys bridges this gap through a two-phase protocol: (i)~semantic
mediation (E1--E2) evaluates the agent's intent against its identity,
input provenance, memory taint, and declared manifests, producing a
dynamically computed allowlist; (ii)~this allowlist is installed as eBPF
probes (E3) before the tool's process tree is spawned, constraining the
actual syscalls that can be issued.  The semantic layer contributes
\emph{intent-awareness} (why is this action being taken?); the kernel
layer contributes \emph{non-bypassability} (the tool cannot escape the
allowlist even if its implementation is malicious).

We state the resulting guarantee:

\begin{property}[Semantic--Syscall Closure]
\label{prop:bridge}
For any tool invocation $t$ with declared manifest $M_t$ and agent
capability boundary $\Smax$, the set of syscalls issued by $t$ and its
descendants is constrained to
$\text{Allow}(t) = \text{SemanticEval}(t, \Smax, M_t, \text{taint})$.
No process in $t$'s tree can issue a syscall outside $\text{Allow}(t)$,
regardless of $t$'s implementation.
\end{property}

This guarantee closes the gap identified in \S\ref{sec:bg:os-needs} and
is, to our knowledge, unique among agent security architectures.
Existing systems address one side or the other: sandboxes provide
non-bypassability without intent-awareness; guardrails provide
intent-awareness without non-bypassability.  \sys's two-phase protocol is
the architectural mechanism that composes both properties.

\subsection{Residual Risks and Limitations}
\label{sec:analysis:residual}

A credible security analysis must acknowledge what remains unprotected.
We enumerate the residual risks that fall outside \sys's current
guarantees.

\parab{Taint fidelity under adversarial segmentation}
The soundness of item-level taint (C1) depends on correct source
segmentation in the cognition pillar.  Adversarially crafted content
designed to blur the boundary between trusted and untrusted segments can
cause the lattice search to compute labels that under-approximate true
taint.  Mitigating this requires robust segmentation techniques, which
remain an open research problem (\S\ref{sec:discussion}).

\parab{Semantic firewall accuracy}
Perception Layer~P3 (Semantic Firewall) is an LLM-based classifier and
inherently trades false positives against false negatives.  An adversary
who understands the classifier's decision boundary can craft payloads in
the blind spot.  The graduated pipeline (P1--P4) reduces this risk
through defense-in-depth but does not eliminate it.

\parab{GAR as a single point of trust}
The GAR is the root of trust for agent identity; its signing key
compromise would enable arbitrary AIC forgery and collapse the identity
pillar.  Cross-GAR federation (\S\ref{sec:discussion}) would distribute
this trust but introduces its own cross-certification challenges.

\parab{Non-bypass assumption}
The entire security argument rests on the assumption that the three
adapter interfaces are the only paths through which agents can reach
LLMs, tools, and storage (Principle~5).  If an agent retains ambient
credentials, direct filesystem access, or alternative tool transports
outside the adapters, mediation is incomplete.  Enforcement of this
assumption is a deployment responsibility.

\parab{eBPF portability}
Execution invariant~E2 depends on eBPF, which is natively supported on
Linux but has limited availability on macOS and Windows
(\S\ref{sec:discussion}).  On non-Linux platforms, alternative backends
must provide equivalent non-bypassability guarantees.

\parab{Formal verification deferred}
The invariants stated in \S\ref{sec:analysis:invariants} are enforced by
construction but have not been mechanically verified.  Formal proofs
(e.g., in Coq or TLA+) of the monotonic attenuation chain and the policy
intersection framework would elevate these guarantees from architectural
claims to machine-checked theorems---work we have deferred to companion
publications (\S\ref{sec:discussion}).

These limitations do not invalidate \sys's architectural contributions,
but they bound the confidence with which the stated guarantees should be
interpreted.  Each limitation corresponds to an active direction in our
research agenda.

%% file: sections/harness.tex
\section{Related Work: From Agent Harness to Agent OS}
\label{sec:harness}

The term ``agent harness'' has rapidly gained currency as the
infrastructure concept for managing the full lifecycle of AI
agents---encompassing deployment, monitoring, guardrails, tool
orchestration, and multi-agent coordination. A growing ecosystem of
projects addresses different slices of this harness. While these efforts
are individually valuable, we argue that they collectively reveal a
deeper need: agents require not just a stronger harness but an
\emph{operating system}---a mandatory, non-bypassable mediation layer
that provides identity, perception, cognition, and execution services as
integrated kernel functions.

This section is organized as follows.
We first present Table~\ref{tab:comparison}, which compares six
representative systems---drawn from the runtime, governance, and sandbox
tiers---against \sys across 15~dimensions that map directly to the four
architectural pillars of \S\ref{sec:architecture}, followed by a
dimension-by-dimension methodology that defines each row and justifies
every non-trivial rating.
\S\ref{sec:harness:landscape} then surveys the broader harness landscape
across four tiers: orchestration frameworks
(\S\ref{sec:harness:landscape}), agent runtimes
(\S\ref{sec:related:runtimes}), governance platforms
(\S\ref{sec:related:governance}), and execution sandboxes---providing
the detailed evidence behind the table's ratings.
Orchestration frameworks (e.g., LangChain, AutoGen) are omitted from
the table because they do not claim OS-level services along these
dimensions; their complementary role is discussed in
\S\ref{sec:harness:landscape}.
\S\ref{sec:harness:gap} distills the cross-tier pattern into the
argument for a missing OS layer.
\S\ref{sec:harness:tcb} positions \sys as the agent OS kernel.
\S\ref{sec:harness:a2a} discusses implications for agent-to-agent and
human-to-agent interactions, and \S\ref{sec:harness:other} covers
additional related areas.

\begin{table*}[t]
  \centering
  \caption{%
    Comparison of representative systems across systematic Agent OS dimensions.
    Rather than treating security and capability as isolated features, \sys
    demonstrates how OS-level security primitives (e.g., identity, taint tracking)
    directly enable advanced agent capabilities (e.g., cross-org A2A, provenance-aware retrieval).
    $\bullet$~=~comprehensive support as a core design element;
    $\circ$~=~partial or optional support;
    ---~=~absent or not a design focus.%
  }
  \label{tab:comparison}
  \footnotesize
  \setlength{\tabcolsep}{3pt}%
  \begin{tabular}{@{}l ccccccc@{}}
    \toprule
    \textbf{Dimension} &
      \rotatebox{70}{\textbf{\sys}} &
      \rotatebox{70}{\shortstack{AGT\\\scriptsize\cite{microsoft2025governance}}} &
      \rotatebox{70}{\shortstack{AIOS\\\scriptsize\cite{mei2024aios}}} &
      \rotatebox{70}{\shortstack{OpenFang\\\scriptsize\cite{openfang2026}}} &
      \rotatebox{70}{\shortstack{SmythOS\\\scriptsize\cite{smythos2026}}} &
      \rotatebox{70}{\shortstack{Letta\\\scriptsize\cite{letta2026}}} &
      \rotatebox{70}{\shortstack{nono\\\scriptsize\cite{nono2026}}} \\
    \midrule
    \multicolumn{8}{@{}l}{\scriptsize\textbf{Pillar~1: Identity} (\S\ref{sec:arch:identity})} \\[2pt]
    Cryptographic workload identity
      & $\bullet$ & $\bullet$ & --- & $\circ$ & --- & --- & $\circ$ \\
    Multi-dimensional provenance
      & $\bullet$ & ---       & --- & --- & --- & --- & --- \\
    Cross-org A2A attestation
      & $\bullet$ & $\circ$   & --- & --- & --- & --- & --- \\
    \midrule
    \multicolumn{8}{@{}l}{\scriptsize\textbf{Pillar~2: Perception} (\S\ref{sec:arch:perception})} \\[2pt]
    Input provenance tagging
      & $\bullet$ & ---       & --- & $\circ$ & --- & --- & --- \\
    Graduated input filtering
      & $\bullet$ & $\circ$   & --- & $\circ$ & --- & --- & --- \\
    \midrule
    \multicolumn{8}{@{}l}{\scriptsize\textbf{Pillar~3: Cognition} (\S\ref{sec:arch:cognition})} \\[2pt]
    Context \& memory management
      & $\bullet$ & --- & $\circ$ & $\circ$ & $\circ$ & $\bullet$ & --- \\
    Memory taint tracking
      & $\bullet$ & ---       & --- & $\circ$ & --- & --- & --- \\
    Provenance-aware retrieval
      & $\bullet$ & ---       & --- & --- & --- & --- & --- \\
    \midrule
    \multicolumn{8}{@{}l}{\scriptsize\textbf{Pillar~4: Execution} (\S\ref{sec:arch:execution})} \\[2pt]
    Tool \& workflow orchestration
      & $\bullet$ & --- & $\circ$ & $\bullet$ & $\bullet$ & --- & --- \\
    Agent scheduling \& concurrency
      & $\bullet$ & --- & $\bullet$ & $\circ$ & $\circ$ & --- & --- \\
    Syscall-level sandboxing
      & $\bullet$ & ---       & --- & $\circ$ & --- & --- & $\bullet$ \\
    Semantic-to-syscall enforcement
      & $\bullet$ & ---       & --- & --- & --- & --- & --- \\
    Plan--trace alignment
      & $\bullet$ & ---       & --- & $\circ$ & --- & --- & --- \\
    \midrule
    \multicolumn{8}{@{}l}{\scriptsize\textbf{Cross-Pillar} (Principle~3, \S\ref{sec:principles})} \\[2pt]
    Cross-lifecycle integration
      & $\bullet$ & ---       & --- & $\circ$ & --- & --- & --- \\
    \bottomrule
  \end{tabular}
\end{table*}

\paragraph{Comparison methodology.}
Each row in Table~\ref{tab:comparison} corresponds to a concrete
capability defined in the \sys architecture
(\S\ref{sec:architecture}).  We group them by pillar and briefly
define each dimension below.

\textbf{Pillar~1: Identity} (\S\ref{sec:arch:identity}).
\emph{Cryptographic workload identity}: whether agent identity is
cryptographically bound to builder, code artifact, operator, and
deployment context---not merely a self-declared string or API key.
AGT earns~$\bullet$ via AgentMesh's SPIFFE/SVID attestation;
OpenFang earns~$\circ$ for Ed25519-signed manifests that do not bind all
four dimensions; nono earns~$\circ$ for Sigstore bundle signing.
\emph{Multi-dimensional provenance}: whether identity material
traces across developer, code, operator, and runtime simultaneously.
No compared system binds all four dimensions; AGT's SPIFFE-based
workload identity covers the runtime dimension alone (already credited
under cryptographic workload identity above).
\emph{Cross-org A2A attestation}: mutual identity verification
between agents from distinct organizations.  AGT earns~$\circ$ via
AgentMesh's SPIFFE federation; no other system provides this.

\textbf{Pillar~2: Perception} (\S\ref{sec:arch:perception}).
\emph{Input provenance tagging}: whether all inbound content is annotated
with origin metadata (source identity, trust level, timestamp) before
reaching the LLM context, corresponding to Layer~P1 (Source Tagger).
These tags persist through the pipeline and are consumed by downstream
cognition and execution pillars.
OpenFang earns~$\circ$ because its \texttt{TaintSource}/\texttt{TaintLevel}
labels attach origin information to messages within the runtime, but
do not produce the full \emph{ProvenanceTag} metadata (origin, timestamp,
trust level, detection warnings) that \sys's P1 generates; no other
compared system tags inputs with provenance at the perception boundary.
\emph{Graduated input filtering}: a multi-layer perception pipeline
(P1--P4) that applies increasingly expensive defenses in sequence.
AGT earns~$\circ$ because its policy engine evaluates declarative rules
on inter-agent messages before they reach the agent, providing
single-layer input checking but not a graduated multi-layer pipeline;
OpenFang earns~$\circ$ for its prompt-injection scanner, which
operates as a single detection pass without layered depth.

\textbf{Pillar~3: Cognition} (\S\ref{sec:arch:cognition}).
\emph{Context \& memory management}: structured lifecycle management of
agent memory (archival, retrieval, scoping).  Letta earns~$\bullet$ for
its self-managed context windows; AIOS earns~$\circ$ for its
\texttt{StorageManager}/LSFS mount-point-based file isolation;
OpenFang earns~$\circ$ for per-message state tracking implied by its
taint-label infrastructure; SmythOS earns~$\circ$ for its pluggable
Storage/VectorDB connectors.
\emph{Memory taint tracking}: information-flow labels propagated at
per-entry granularity.  OpenFang earns~$\circ$ for its
\texttt{TaintSource}/\texttt{TaintLevel} labels, which operate within
the runtime but do not propagate through a graduated perception pipeline.
\emph{Provenance-aware retrieval}: retrieval decisions conditioned on
taint labels and provenance chains, not just semantic similarity.
No compared system implements this.

\textbf{Pillar~4: Execution} (\S\ref{sec:arch:execution}).
\emph{Tool \& workflow orchestration}: managing tool routing, skill
composition, and multi-step workflows.  OpenFang and SmythOS
earn~$\bullet$ for rich plugin/connector ecosystems; AIOS earns~$\circ$
for its syscall-style tool API.  AGT's policy engine evaluates
governance rules on tool calls but does not perform tool routing or
workflow composition, hence~---.
\emph{Agent scheduling \& concurrency}: concurrent agent execution
with isolation.  AIOS earns~$\bullet$ for its FIFO/round-robin
scheduler; OpenFang and SmythOS earn~$\circ$ for limited concurrency
support.
\emph{Syscall-level sandboxing}: OS-kernel-level process isolation
(eBPF, Landlock, Seatbelt).  nono earns~$\bullet$ for
Landlock/Seatbelt-based sandboxing; OpenFang earns~$\circ$ for WASM
metering sandboxes that do not reach the host kernel.
\emph{Semantic-to-syscall enforcement}: bridging high-level
permission semantics to low-level kernel hooks (Layers E1--E3).
No compared system links semantic policy to syscall-level enforcement.
\emph{Plan--trace alignment}: post-execution verification that
compares declared plans to actual execution traces (Layer~E4).
OpenFang earns~$\circ$ for its Merkle audit chains, which provide
tamper-evident recording of actions but do not automatically align
recorded traces against stated plans.  AGT's pre-execution policy checks
and SLO monitoring are governance mechanisms, not post-hoc plan--trace
comparison, hence~---.

\textbf{Cross-Pillar} (Principle~3, \S\ref{sec:principles}).
\emph{Cross-lifecycle integration}: whether identity, perception,
cognition, and execution share provenance labels and taint metadata
across the full agent lifecycle.  OpenFang earns~$\circ$ because its
security features (taint labels, audit chains, capability gating) compose
within the runtime for hosted agents but do not extend across externally
connected agents.  AGT's components span multiple lifecycle phases but
are independently adoptable modules that do not share unified provenance
or taint metadata, hence~---.

\textbf{Scoring criteria.}
$\bullet$~indicates comprehensive, architecturally integrated support
as a core design element;
$\circ$~indicates partial or optional support (e.g., available as an
add-on, limited to a subset of agents, or lacking cross-pillar
integration);
---~indicates the capability is absent or not a design focus.
Ratings reflect our reading of each project's public documentation and
open-source repositories as of May~2026; they are qualitative author
assessments, not independently audited certifications.

\subsection{The Agent Harness Landscape}
\label{sec:harness:landscape}

\subsubsection{Orchestration Frameworks}
Orchestration frameworks (e.g., LangChain/LangGraph~\cite{langchain2025,langgraph2025},
AutoGen~\cite{autogen2024}, Microsoft Agent Framework~\cite{ms-agent-framework2026},
Semantic Kernel~\cite{semantickernel2025}, CrewAI~\cite{crewai2025},
MetaGPT~\cite{metagpt2024}, AutoGPT~\cite{autogpt2025}, and smolagents~\cite{smolagents2025})
collectively represent widely adopted agent infrastructure, with large downstream
ecosystems spanning vendors, cloud providers, and open-source communities. They manage agent workflows, tool routing,
retries, and inter-agent messaging.

\textbf{Distinction:} Their primary contribution is building \emph{stronger agents}---better
reasoning chains, more reliable tool use, and smoother multi-agent
coordination. Security ranges from absent to optional enterprise plugins (e.g., smolagents documents that its built-in \texttt{LocalPythonExecutor} is not a security sandbox and recommends external isolation for untrusted code~\cite{smolagents2025}). \sys integrates with any orchestrator through three narrow adapter interfaces (Principle~5) and provides the OS layer that no orchestration framework offers. The relationship is complementary: orchestrators build capable agents; \sys provides the OS those agents run on.

\subsubsection{Agent Runtimes}
\label{sec:related:runtimes}

A growing class of projects positions itself between low-level
orchestration frameworks and full governance platforms by providing
\emph{runtime infrastructure services} for AI agents, often borrowing
OS terminology such as ``kernel,'' ``scheduling,'' and ``memory
management.''  We call these \emph{agent runtimes} and note that the
term covers a wide spectrum of deployment models, abstraction levels,
and security ambitions.  To disambiguate this spectrum---and to
clarify how it relates to \sys---we first characterize four
representative systems from their source code and public documentation,
then summarize the structural differences in
Table~\ref{tab:runtime-comparison}.

\paragraph{What agent runtimes provide.}
Despite the shared ``OS'' framing, agent runtimes differ sharply in
\emph{what} they abstract and \emph{how} they enforce:

\begin{itemize}
  \item \textbf{AIOS}~\cite{mei2024aios} (and the conceptual
    ``LLM\,as\,OS'' paper~\cite{ge2023llmos}) ships a user-space
    ``kernel'' as a \emph{FastAPI/uvicorn HTTP server}: the Cerebrum
    agent SDK talks to it via \texttt{POST /query} over localhost, while
    agents submitted through \texttt{/agents/submit} run inside the same
    server process in a thread pool.
    The kernel wraps a request queue and FIFO/round-robin scheduler for
    concurrent agents; a \texttt{StorageManager} (LSFS) offers
    mount-point-based file isolation; and a syscall-style API
    (\texttt{llm\_syscall}, \texttt{memory\_syscall},
    \texttt{tool\_syscall}) defines the interaction surface.
    Despite the server boundary, the separation is an \emph{API
    convention}: agents co-hosted in the server share the same Python
    address space and can bypass the syscall layer directly, while
    remotely connected agents gain no stronger isolation than an
    unauthenticated HTTP endpoint (the API has no authentication
    middleware).  Access control exists as a stub
    \texttt{PermissionManager} without cryptographic identity or
    taint labels.

  \item \textbf{OpenFang}~\cite{openfang2026} compiles to a
    \emph{single Rust binary}: \texttt{openfang start} runs
    \texttt{OpenFangKernel} plus an Axum HTTP API in one daemon
    (default \texttt{127.0.0.1:4200}), while CLI commands without a
    daemon boot the same kernel in-process.  External clients reach
    agents via REST/WebSocket/SSE; kernel-hosted agents execute
    \emph{inside} the daemon (LLM loops as async tasks; optional WASM
    or Python modules) and invoke kernel services through an in-process
    \texttt{KernelHandle} callback rather than HTTP.  Its security
    feature set is the richest among the runtimes: WASM dual-metering
    sandboxes (wasmtime with instruction-count and wall-clock epoch
    limits), Ed25519-signed agent manifests, Merkle audit chains,
    information-flow taint labels (\texttt{TaintSource}/\texttt{TaintLevel}),
    SSRF host allowlists, prompt-injection scanning, capability-gated
    tool access with parent$\to$child inheritance validation, and
    MCP/A2A protocol adapters.  For these hosted agents, mediation is
    mandatory---every inbound message traverses quota checks,
    capability filtering, and audit logging before the agent loop, and
    child agents cannot exceed their parent's capability set.  For
    \emph{externally connected} agents (e.g., a LangChain A2A service),
    the runtime is only an integration endpoint: A2A/MCP forwarding
    does not bring the peer's LLM calls, memory, or tool use inside the
    kernel perimeter.

  \item \textbf{SmythOS SRE}~\cite{smythos2026} brands itself as
    ``The Linux of AI Agents.''  Its open-source Smyth Runtime
    Environment is an \emph{in-process Node.js library}: a singleton
    \texttt{SRE.init()} bootstraps 16 subsystem services (LLM, Storage,
    VectorDB, Vault, Scheduler, Telemetry, etc.)\ inside the host
    application's process.  The ``OS'' metaphor materializes as a
    \emph{Connector pattern}: pluggable backends (local/S3, RAM/Redis,
    Pinecone, etc.)\ expose a unified API so that
    \texttt{storage.write()} is backend-agnostic.  Each agent receives
    a scoped resource view via an \texttt{AccessCandidate} identity
    (User/Agent/Team), with decorator-enforced ACLs (Read/Write/Owner)
    and SES-based JavaScript sandboxing for user-supplied code.
    However, the runtime carries no cryptographic workload identity,
    taint labels, or kernel-level enforcement boundary; the commercial
    SmythOS cloud platform (separate from this open-source runtime)
    adds hosted infrastructure but not OS-kernel-class security.

  \item \textbf{Letta}~\cite{letta2026} (formerly MemGPT) runs as a
    \emph{standalone FastAPI server} offering ``agents as a service'':
    agents are persistent, stateful objects whose core, archival, and
    recall memory is managed by the server and accessed through
    REST/SDK.  Its core innovation is \emph{self-managed context
    windows}: agents use built-in tools to archive, retrieve, and
    edit their own memory blocks.  Tool sandboxing exists as a
    three-mode option (\texttt{local} / \texttt{e2b} /
    \texttt{local\_sandbox}), defaulting to in-process execution with no
    isolation.  No information-flow labels, provenance tags, or
    cryptographic identities are applied to memory items.
\end{itemize}

Other works in this space include KAOS~\cite{zhuo2025kaos}, which builds a
multi-agent OS on Kylin Linux; AutoAgent~\cite{tang2025autoagent}, which
frames itself as an OS enabling zero-code agent creation; and the
Agent-Centric OS review~\cite{jia2024acos} and AI-in-OS
survey~\cite{zhang2024aiinsideos}, which provide conceptual roadmaps.

\begin{table}[t]
  \centering
  \caption{%
    Agent runtime deployment models and enforcement characteristics.
    \emph{Deployment} indicates whether the runtime runs in the same
    process as the agent (\textbf{in-proc}), as a separate service
    (\textbf{service}), or in a hybrid mode.
    \emph{Security enforcement} indicates whether security mechanisms
    are mandatory or opt-in.
    \sys is included for comparison; its deployment form is explicitly
    secondary to the non-bypass invariant (Principle~5).%
  }
  \label{tab:runtime-comparison}
  \footnotesize
  \setlength{\tabcolsep}{3pt}%
  \begin{tabular}{@{}l l l l@{}}
    \toprule
    \textbf{System} & \textbf{Deployment} & \textbf{Primary focus} & \textbf{Security} \\
    \midrule
    AIOS        & HTTP server + in-proc agents  & LLM scheduling     & No auth; stub ACL \\
    OpenFang    & Daemon / in-proc kernel       & Skill/Plugin sandbox    & Mandatory for hosted; opt-in for external \\
    SmythOS     & In-proc library (npm)         & Unified resources   & ACL + SES sandbox \\
    Letta       & Standalone server             & Memory management   & Optional sandbox \\
    \midrule
    \sys        & Form-agnostic                 & Security kernel     & Mandatory, non-bypassable \\
    \bottomrule
  \end{tabular}
\end{table}

\paragraph{How runtimes relate to agents and to \sys.}
Agent runtimes span a spectrum between ``part of the agent'' and
``mandatory management layer,'' and the distinction matters for security.
An in-process runtime like SmythOS SRE shares the agent's address space;
its ACL decorators can be bypassed by any code that reaches the
underlying connector object.  AIOS runs agents in the same server
process via a thread pool, so co-hosted agents can access kernel
internals directly despite the HTTP syscall facade; remotely connected
agents gain no stronger isolation than an unauthenticated HTTP endpoint.
Even OpenFang---which enforces mandatory capability checks and audit
logging for agents \emph{hosted inside} its kernel---cannot extend those
guarantees to externally connected A2A agents whose LLM calls, memory
accesses, and tool invocations remain outside the kernel's perimeter.
Letta's server-mode boundary isolates agent state but carries no
information-flow constraints on memory items.

\sys differs on a single architectural invariant: the three adapter
interfaces (LLM, tool, storage) constitute the \emph{only} path to the
capabilities those adapters guard (Principle~5).  Whether \sys is
deployed as an embedded SDK, a gateway, or a sidecar is a deployment
decision; the non-bypass guarantee is preserved regardless, because
ambient credentials and direct transports are not available to the
agent.

The deployment similarities between service-mode runtimes and a
sidecar \sys are architectural surface only.  A runtime provides
\emph{opt-in platform services} that agents may or may not use;
\sys provides a \emph{mandatory mediation layer} that agents cannot
circumvent.  Even OpenFang's security modules---while individually
sound---do not compose into a unified lifecycle discipline: taint
labels are tracked within the runtime but do not propagate through
a graduated perception pipeline or govern long-term memory
retrieval, and the enforcement scope is bounded to the
OpenFang-hosted subset of the agent ecosystem.

\textbf{Distinction:}
Agent runtimes advance \emph{agent capability and developer
experience}: better scheduling (AIOS), richer memory (Letta), broader
integrations (SmythOS), and execution isolation (OpenFang).
\sys advances \emph{mandatory, lifecycle-spanning security}: it
provides the non-bypassable OS kernel that these runtimes---or the
orchestrators and governance platforms above and below them---can adopt
to turn their opt-in services into externally auditable guarantees.

\subsubsection{Governance Platforms}
\label{sec:related:governance}

Unlike agent runtimes, which provide the execution substrate itself,
\emph{governance platforms} sit alongside agent code and supply
observability, policy enforcement, and compliance reporting.  Critically,
these platforms integrate via \textbf{code-level SDKs}---imported as
libraries or decorators into the agent's own process---rather than as
external, non-bypassable enforcement boundaries.  This design gives
governance platforms broad framework compatibility but limits their
security guarantees: any code running in the same process can, in
principle, bypass the governance hooks.

\paragraph{Observability and evaluation stacks.}
AgentOps~\cite{dong2024agentops} and LangSmith~\cite{langsmith2025}
provide Python/TypeScript SDKs that instrument agent code through
decorators (\texttt{@traceable}), client wrappers
(\texttt{wrap\_openai}, \texttt{wrap\_anthropic}), and ASGI middleware,
forwarding trace spans, evaluation metrics, and run logs to a hosted
analytics backend.  LangSmith additionally ships an experimental
cloud-hosted sandbox module (Docker-based isolated containers accessed
through a \texttt{SandboxClient} API) for executing untrusted code in a
remote environment---architecturally similar to E2B rather than to an
in-process policy engine.  Despite this execution capability, the
governance relationship remains \emph{opt-in}: agents may bypass the
instrumented path or invoke tools outside the sandbox, because no
mandatory interposition prevents direct access to LLMs, local tools, or
memory.

\paragraph{Policy-enforcement toolkits.}
Microsoft's Agent Governance Toolkit~(AGT)~\cite{microsoft2025governance}
is the most comprehensive industry effort.  It ships as a set of
\emph{application-level Python packages} that run \emph{inside the same
process} as the agent application.  Its components include:
\begin{itemize}
  \item \textbf{Agent OS} (confusingly named---an in-process
    deterministic policy engine, not an OS kernel): evaluates
    declarative policies before each tool call or inter-agent message;
  \item \textbf{AgentMesh}~\cite{agentmesh}: a sidecar-deployable
    zero-trust layer providing SPIFFE/SVID workload identity and
    protocol bridges (A2A, MCP, IATP);
  \item \textbf{Agent Hypervisor}: execution rings with resource
    quotas and circuit-breaker semantics;
  \item \textbf{Agent SRE}: SLO monitoring, chaos injection, and
    automated remediation.
\end{itemize}
Because the core policy engine is an in-process library, agents and the
governance logic share a single trust boundary.  A compromised agent---or
any code with access to the interpreter---can mutate policy state,
skip evaluation hooks, or invoke tools directly.  AgentMesh provides genuine cryptographic workload identity via
out-of-process SPIFFE/SVID attestation (hence~$\bullet$ for that row in
Table~\ref{tab:comparison}) and enables cross-org attestation through
SPIFFE federation ($\circ$).  However, SPIFFE binds the runtime workload
but does not trace provenance across developer, code artifact, and
operator simultaneously (hence~--- for multi-dimensional provenance),
and tool execution and memory access still traverse no mandatory
mediation.

Other work explores agent identity management~\cite{south2025identitymanagement},
zero-trust authentication frameworks~\cite{zt-agentic-ai}, and
cryptographic human-delegation provenance~\cite{hdp}.

\paragraph{Distinction:}
AGT and \sys share the \emph{goal} of governing agent actions but differ
on a fundamental architectural choice: \emph{where} the governance logic
executes relative to the agent's trust boundary.
AGT interposes governance as a co-located library---modular,
independently adoptable, and compatible with any Python/TS/.NET/Rust/Go
agent~\cite{microsoft2025governance}---but cannot guarantee
non-bypassability because governance and agent share an address space.
\sys places governance in a \emph{separate kernel} that controls the
\emph{only} interfaces through which agents reach LLMs, tools, and
storage (Principle~5).  Identity, perception, cognition, and execution
share unified taint metadata enforced below the agent's address space
(e.g., via eBPF-class hooks), making bypass equivalent to escaping the
OS kernel.
Where AgentMesh adds SPIFFE/SVID identity as an adjacent sidecar, \sys
folds mutual attestation, verifiable credentials, and delegation
semantics into the same kernel abstraction that mediates tool and
storage access---eliminating the gap between ``who you are'' and ``what
you may do.''
The relationship is thus complementary at different trust levels: AGT's
rich policy language and framework adapters can run \emph{above} \sys,
gaining non-bypassable enforcement from the kernel below without
sacrificing their application-level expressiveness.

\subsubsection{Execution Sandboxes}
Execution sandboxes isolate agent execution at the OS or cloud level.
nono~\cite{nono2026} offers Landlock/Seatbelt-based kernel sandboxing,
Sigstore-based signing of agent instruction bundles, credential proxy injection, and network allowlists.
E2B~\cite{e2b2026} provides cloud-side isolated environments with
Python/JS SDKs. Anthropic's sandbox-runtime~\cite{anthropic-sandbox2026} enforces
filesystem and network restrictions at the OS level without a container runtime.

\textbf{Distinction:} These tools provide effective \emph{execution-layer isolation}.
They do not, by themselves, subsume \sys's full-stack treatment: even where a
sandbox adds supply-chain controls (e.g., Sigstore attestation of skill
instruction files in nono~\cite{nono2026}), it does not replace graduated
perception filtering and memory taint propagation across the entire agent
pipeline. Their isolation is primarily binary (inside vs.\ outside the
sandbox), whereas \sys provides \emph{graduated, context-aware} enforcement.
Each tool invocation receives a dynamically computed allowlist derived from
the agent's identity, the provenance of its inputs, and the taint level of the
memory entries. \sys's execution pillar subsumes sandbox functionality and
extends it with semantic context.

\subsection{The Missing OS Layer: Why an OS, not just a stronger harness?}
\label{sec:harness:gap}

The pattern across these four tiers is revealing. Each tier excels within its scope, but no tier provides a \emph{unified} OS layer with identity, perception, cognition, and execution as integrated services sharing provenance labels, trust signals, and taint metadata. The gap is not a missing feature; it is a missing \emph{architectural layer}.

The distinction between an agent harness and an agent OS parallels the
distinction between a user-space library and an OS kernel. A library
provides services to applications that \emph{choose} to call it;
a kernel provides mandatory services that \emph{all} applications must
traverse. Current harness solutions operate as opt-in libraries: an
agent framework can skip the guardrail, bypass the policy engine, or
ignore the sandbox. \sys operates as a mandatory kernel: because it
controls the \emph{only} interfaces through which agents can reach LLMs,
tools, and storage (Principle~5), a compromised or misconfigured agent cannot bypass the OS layer any more than a compromised user-space process can bypass the OS kernel.

This mandatory mediation simultaneously improves both security and
capability. Security improves because enforcement is non-bypassable.
Capability improves because operators can \emph{trust} the enforcement
boundary, enabling them to grant agents broader tool access, richer
memory sharing, and deeper delegation chains---all of which would be
too risky without OS-level guarantees.

\subsection{\sys as the Agent OS Kernel}
\label{sec:harness:tcb}

\sys provides the \emph{operating system kernel} that transforms
a collection of harness components into an integrated, trustworthy
agent platform. The kernel consists of three components: the \emph{Global Agent
Registry} (the remote trust anchor), the \emph{Agent Kernel} (the local
trust anchor holding the private key), and the \emph{three adapter
interfaces} (LLM adapter, tool adapter, storage adapter) through which
all capability interactions must pass. Everything else lies \emph{outside} the
kernel and is assumed potentially compromised.

The kernel contributes three services that no orchestration or
governance layer alone provides:
\begin{enumerate}
  \item \textbf{Identity infrastructure:} Mutual attestation, verifiable AICs, and capability-bounded delegations.
  \item \textbf{Information-flow control across the pipeline:} Provenance labels flow from perception through cognition to execution.
  \item \textbf{Execution governance with kernel-level enforcement:} Tool executions exceeding declared permissions are blocked at the OS level, not just logged.
\end{enumerate}

\sys inverts the relationship where infrastructure logic is entangled with agent business logic. The OS kernel is a \emph{standalone component} that can integrate as an \emph{embedded SDK}, \emph{middleware}, or a \emph{sidecar}. The agent framework can be swapped without re-implementing OS services.

\subsection{Implications for A2A and H2A Interactions}
\label{sec:harness:a2a}

\paragraph{Agent-to-Agent (A2A).}
\sys's identity pillar provides the necessary trust infrastructure for agents from different organizations to interoperate~\cite{blocka2a,google2025a2a}. Mutual attestation establishes identities, and session tokens scope the interaction.

\paragraph{Human-to-Agent (H2A).}
Users need assurance that the agent acted as instructed. \sys's execution pillar provides this through plan--trace alignment and immutable audit logs, while the OIDC-bound operator identity ensures accountability.

\subsection{Other Related Areas}
\label{sec:harness:other}

\paragraph{GUI and computer-use agents.}
A vibrant line of work builds agents that interact through graphical
user interfaces~\cite{wu2024oscopilot,agashe2024agents,zhang2025ufo,%
ye2025mobileagentv3,wu2024osatlas,bonatti2024windowsarena}, with
comprehensive surveys~\cite{hu2025osagents,tang2025guisurvey} and
safety benchmarks including OSWorld~\cite{xie2024osworld},
OS-Harm~\cite{kuntz2025osharm}, and PASB~\cite{wang2026openclaw}.
These focus on \emph{interaction modality} and operate \emph{within}
the OS as applications; \sys operates \emph{below} such agents as
infrastructure and can serve as the enforcement layer.

\paragraph{Agent memory systems.}
MemOS~\cite{li2025memos} and MemoryOS~\cite{kang2025memoryos} optimize memory
\emph{capacity and retrieval quality}; their published designs do not center on
memory \emph{security} in the sense of information-flow labels or taint
propagation across agent subsystems. \sys's cognition pillar supplies that
missing enforcement-oriented layer.

\paragraph{Multi-agent platforms.}
Platforms like OpenHands~\cite{wang2024openhands}, SWE-agent~\cite{yang2024sweagent}, Magentic-One~\cite{fourney2024magneticone}, and AgentStore~\cite{jia2025agentstore} are engineering platforms where security is optional; \sys provides the OS kernel that such platforms can adopt.

\paragraph{Agent protocols.}
Protocols like MCP~\cite{anthropic2024mcp}, A2A~\cite{google2025a2a}, ANP~\cite{anp-whitepaper}, Agent-OSI~\cite{agent-osi}, and BlockA2A~\cite{blocka2a} specify \emph{what agents can say}; \sys ensures \emph{what agents can do}---the two are complementary.

%% file: sections/discussion.tex
\section{Discussion and Future Directions}
\label{sec:discussion}

Beyond the design and analysis presented above, we summarize several architectural insights, examine their implications for how the field
thinks about agent security, and outline the open problems and research
directions they motivate.

\subsection{Architectural Insights}
\label{sec:discussion:insights}

\parab{The agent is outside the TCB}
A conventional reading of agent security treats the agent---its reasoning
loop, its tools, its memory---as the entity to be \emph{protected}.  \sys's
architecture inverts this: the agent (LLM, application logic, orchestration
fabric) is treated as an \emph{untrusted compute substrate} that the security
kernel \emph{contains}.  The kernel mediates every interaction---inbound
perception, memory storage and retrieval, tool invocation---and the agent
cannot reach any resource without kernel approval.  This inversion
generalizes beyond \sys: it implies that agent security architectures should
not ask ``how do we protect the agent?'' but ``which components must we trust
to constrain an agent that may behave adversarially?''  Concretely, the
trusted computing base consists of the GAR, the Agent Kernel, the three
adapters, the eBPF subsystem, and the cryptographic primitives
(\S\ref{sec:analysis:threat}).  The agent, its LLM, and its application code
are explicitly excluded.

\parab{Provenance as the security nervous system}
A second insight is that provenance is not merely a feature of one pillar but
the \emph{coupling substrate} that transforms four independent enforcement
mechanisms into a single coherent security argument.  Identity establishes
\emph{who} produced content; Perception assigns initial trust levels;
Cognition propagates taint labels through memory derivations; Execution
consumes these labels to build allowlists and constrain tool actions.  This
provenance flow---labeled, auditable, and mandatory---is the agent-level
analogue of labeling in classical mandatory access control (Bell--LaPadula,
Biba)~\cite{bell1973secure,biba1977integrity}, but applied to the semantic
plane rather than to files and processes.  The architectural lesson is that
provenance must be a \emph{first-class OS primitive} in agent systems, not an
application-level annotation that agents may or may not maintain.

\parab{The reference monitor at the semantic plane}
Classical OS kernels mediate physical resources: memory pages, file
descriptors, process address spaces.  \sys demonstrates that the same
architectural pattern---a non-bypassable, tamper-proof, always-on authority
that mediates every resource transition~\cite{anderson1972planning}---applies
to \emph{semantic resources}: identity claims, perception labels, memory
taint, and execution intent.  This suggests that the reference monitor
concept is more general than its conventional instantiation in OS kernels:
when the resources to be mediated are semantic (natural language, memory
entries, tool plans), the reference monitor operates at the semantic level
while anchoring enforcement to the physical syscall boundary.  The
semantic--syscall bridge (Property~\ref{prop:bridge}) is the architectural
mechanism that realizes this dual-level mediation.

\parab{Scope of the lifecycle decomposition}
The four-pillar decomposition follows from the perception--cognition--execution
lifecycle with identity cutting across all phases (\S\ref{sec:bg:lifecycle}).
This decomposition is appropriate for \emph{interactive, tool-augmented, memory-bearing}
agents---the dominant paradigm in current practice.  Purely reactive agents,
agent swarms with emergent rather than hierarchical delegation, and agents
that do not maintain persistent memory may not require all four pillars.  In
such settings, \sys degrades gracefully: the identity pillar operates
independently; the perception and execution pillars remain mandatory for any
agent that ingests external content or invokes tools; only the cognition
pillar's information-flow guarantees depend on the presence of durable
memory.  Characterizing the minimal pillar configuration for a given agent
class is an open question.

\subsection{Security as a Capability Multiplier}
\label{sec:discussion:capability}

A recurrent concern with security-first designs is that security constrains
what agents can do.  \sys's architecture demonstrates the opposite: each
security pillar simultaneously \emph{enables} agent capabilities that are
unattainable without OS-level enforcement.  This is not a coincidental
property but a structural consequence of non-bypassable mediation.

The underlying principle can be stated precisely: in any agent system with
non-bypassable security mediation, the set of capabilities that can be
\emph{safely delegated} to the agent grows monotonically with the strength of
the security invariants.  The reason is that non-bypassable enforcement
transforms the operator's question from ``can the agent be trusted with
action X?'' to ``can the agent be \emph{prevented} from exceeding the
permitted scope of X?'' ---a fundamentally easier question because the answer
depends on the correctness of the enforcement layer, not on the agent's
future behavior.

This logic plays out concretely across the four pillars:

\begin{itemize}
  \item \textbf{Identity $\to$ cross-organizational A2A.}
    Without cryptographically verified identity and kernel-mediated mutual
    attestation, agent-to-agent collaboration is confined to agents within a
    single trust domain, because verifying the counterparty's identity,
    authorization, and delegation chain is infeasible at the application
    layer.  \sys's identity pillar---where $\Kpriv$ never leaves the kernel
    boundary and the GAR serves as the root of trust---unlocks
    marketplace-style agent discovery, cross-organization delegation, and
    accountable multi-hop workflows.

  \item \textbf{Perception $\to$ richer input channels.}
    When all external content enters the LLM context without mandatory
    mediation, operators rationally restrict input modalities as a
    precaution---disabling browser access, stripping multimodal content, or
    requiring human review at every turn.  The graduated perception pipeline
    (P1--P4) provides defense-in-depth with provenance-tagged outputs,
    enabling agents to safely ingest richer and higher-entropy input sources.

  \item \textbf{Cognition $\to$ trustworthy long-horizon memory.}
    Without provenance labels, retrieval from long-term memory degrades as
    stores grow: all entries appear equally trustworthy, and a single
    poisoned entry can silently influence decisions across sessions.
    \sys's item-level taint propagation enables provenance-weighted
    retrieval, so high-confidence entries are preferentially surfaced and
    contaminated entries carry explicit warnings---turning memory from a
    liability into a reliable asset for long-horizon reasoning.

  \item \textbf{Execution $\to$ expanded tool access with non-bypassable
    guardrails.}  Application-layer guardrails can inspect tool
    \emph{parameters} but cannot constrain tool \emph{execution}: a malicious
    or buggy tool can spawn child processes, access undeclared files, or open
    network connections.  The semantic--syscall bridge (E1--E3) translates
    declared permissions into dynamically installed eBPF allowlists before
    the tool's process tree is spawned, giving operators confidence to expand
    tool sets and grant write access that would be too risky with
    application-layer-only enforcement.
\end{itemize}

The net effect is that \sys does not trade security against capability.
Security is the \emph{mechanism} through which capability is safely expanded:
stronger invariants at each pillar translate directly into broader, safer
delegation.  Returning to the motivating DevOps scenario
(\S\ref{sec:intro:scenario}), each failure in the chain---impersonation,
injection, memory poisoning, over-privileged execution---corresponds to an
absent invariant; supplying that invariant simultaneously closes the security
gap and removes the corresponding capability ceiling.

\subsection{Integration with the Agent Ecosystem}
\label{sec:discussion:integration}

\sys is designed as a complementary OS layer, not a replacement for existing
agent infrastructure.  This design choice reflects a classical systems
principle: different layers of the stack optimize different concerns, and
clean interfaces enable independent evolution.  Just as TCP/IP did not
replace Ethernet and the web did not replace TCP/IP, \sys adds a
\emph{mandatory security mediation layer} beneath orchestration frameworks,
agent runtimes, governance platforms, and execution sandboxes---each of which
continues to optimize its own concern (workflow composition, scheduling,
policy authoring, process isolation).

The three adapter interfaces (LLM, tool, storage) define the integration
boundary.  Any orchestration framework or agent runtime can adopt \sys by
routing its model calls, tool invocations, and memory operations through
these adapters, gaining identity, perception, and taint-aware memory without
re-architecting its scheduling or workflow logic:

\begin{itemize}
  \item \textbf{Beneath agent runtimes.}  AIOS~\cite{mei2024aios} or
    OpenFang~\cite{openfang2026} can retain their scheduler and resource
    manager while \sys provides the security substrate: identity-driven
    workload isolation, perception mediation for all LLM-bound content, and
    provenance-aware memory for cross-session state.

  \item \textbf{Beneath governance platforms.}  Microsoft's Agent Governance
    Toolkit (AGT)~\cite{microsoft2025governance} provides policy authoring
    and audit pipelines; \sys provides the non-bypassable enforcement
    substrate beneath AGT's policy engine, closing the gap between policy
    evaluation and kernel-level constraint installation.

  \item \textbf{Above execution sandboxes.}  Execution sandboxes such as
    nono~\cite{nono2026} and E2B~\cite{e2b2026} provide process isolation;
    \sys provides the semantic context---identity, taint, intent
    alignment---that determines \emph{what} the sandbox should permit for a
    given invocation.  The sandbox enforces isolation; \sys supplies the
    dynamically computed policy.
\end{itemize}

This layered model is not merely a pragmatic accommodation of existing
ecosystem investments.  It is an architectural commitment: the agent OS
kernel should be \emph{minimal}---it should mediate only what must be
enforced below the agent's control plane---and delegate to higher or lower
layers for concerns best addressed there.  The three adapters are the
contract that bounds this minimality.

\subsection{Limitations and Open Problems}
\label{sec:discussion:limitations}

Section~\ref{sec:analysis:residual} analyzed residual risks that fall outside
\sys's current security guarantees: taint fidelity under adversarial
segmentation, semantic firewall accuracy, GAR centralization, the non-bypass
assumption, eBPF portability, and deferred formal verification.  Here we
address broader architectural and systems challenges that constrain
\sys's deployability and motivate the research agenda in
\S\ref{sec:discussion:agenda}.

\parab{LLM-based components in the TCB}
Perception Layer~P3 (Semantic Firewall) and Execution Layer~E2 (LLM
Validator) are LLM-based classifiers that sit \emph{inside} the trusted
computing base: their decisions directly affect which content reaches the
agent and which actions are permitted.  The adversarial robustness of these
components is not yet characterized for agent-security workloads, and the
defense-in-depth argument (Principle~3) mitigates but does not eliminate the
risk of a classifier-level failure cascading across pillars.  Reducing
reliance on LLM-based enforcement---or proving that the remaining
deterministic layers (P1--P2, E1, E3) provide sufficient independent
backstop---is a priority for strengthening the TCB.

\parab{TCB size and heterogeneity}
The combined TCB (GAR, Agent Kernel, three adapters, eBPF subsystem, LLM
classifiers, and cryptographic primitives) is large and spans multiple
implementation domains (distributed services, kernel modules, language-model
inference).  While the adapters narrow the agent-visible attack surface
(Principle~5), the TCB \emph{itself} has not been minimized.  A natural
direction is to ask which components are strictly necessary: can the LLM
Validator (E2) be eliminated if deterministic checks (E1) and eBPF
enforcement (E3) are sufficiently expressive?  Can the GAR be decentralized
through threshold signatures or a transparency log, reducing the trust
concentrated in a single CA?

\parab{Policy specification and human factors}
\sys's security guarantees assume correct policy---capability boundaries
$\Smax$, tool manifests, taint lattice configurations---but the paper has
not addressed the human problem of \emph{specifying} these policies.  Who
writes an agent's capability boundary?  How are tool manifests validated
against implementation behavior?  When policies from different organizations
conflict in a cross-GAR deployment, what is the conflict resolution
procedure?  Operator-facing tooling for policy authoring, simulation, and
audit is essential for deployment feasibility and represents a significant
systems and HCI research challenge.

\parab{Non-bypassability in heterogeneous deployments}
The guarantees in \S\ref{sec:analysis:invariants} are predicated on
Principle~5: the three adapters must be the \emph{only} paths through which
agents reach LLMs, tools, and storage.  In heterogeneous production
environments---where agents may have pre-existing ambient credentials, direct
filesystem access, or side channels through orchestration
infrastructure---ensuring this property is an operational challenge, not a
theoretical one.  Automated verification of deployment integrity (e.g.,
checking that no file descriptors or network sockets bypass the adapters)
would strengthen the deployment-time security argument.

\parab{Empirical characterization}
While the architecture targets low enforcement overhead, comprehensive
performance and security characterization under realistic workloads is
pending.  Three empirical questions are particularly salient: (i)~the latency
distribution of eBPF probe installation/teardown per tool invocation under
concurrent agent workloads; (ii)~the scalability of lattice-search operations
over large memory stores with millions of provenance-tagged entries; and
(iii)~the end-to-end security posture under adversarial stress, including
taint fidelity when source-segmentation boundaries are adversarially blurred
(\S\ref{sec:analysis:residual}) and semantic firewall calibration across
diverse application domains.

\subsection{Research Agenda}
\label{sec:discussion:agenda}

The limitations above define a research agenda organized around four themes.

\parab{Formal foundations}
Mechanized verification of the structural guarantees claimed in
Sections~\ref{sec:analysis:invariants} and~\ref{sec:analysis:structural}
would elevate them from architectural claims to machine-checked theorems.
Priority targets include: the monotonic attenuation chain
(Property~\ref{prop:monotonic}), verified via Coq or Lean against the AIC
issuance and kernel-mediated delegation protocols; the policy intersection
semantics (Principle~2), formalized as a lattice of capability sets with
intersection as the composition operator; and TLA+ models of the mutual
attestation and delegation protocols (\S\ref{sec:arch:identity}), covering
liveness under concurrent delegation and safety under GAR compromise.
Formalizing the semantic--syscall bridge (Property~\ref{prop:bridge}) would
connect these semantic-level models to the operational semantics of eBPF
enforcement.

\parab{Empirical security evaluation}
The companion empirical program should go beyond performance benchmarking to
include adversarial evaluation of each pillar and of cross-pillar
composition.  For Perception: measuring false-positive/false-negative rates
of the graduated pipeline against established prompt-injection and jailbreak
benchmarks (OS-Harm~\cite{kuntz2025osharm}, PASB, MobileSafetyBench).  For
Cognition: quantifying taint fidelity degradation under adversarially crafted
content designed to blur source-segment boundaries.  For Execution:
evaluating whether plan--trace alignment (E4) reliably surfaces
hallucinated or unauthorized actions, and whether the eBPF allowlist
construction (E3) introduces race conditions under concurrent tool
invocations.  For composition: red-team exercises that attempt to exploit
cross-pillar coupling (e.g., crafting a payload that simultaneously evades
perception and induces incorrect taint labels that propagate to permissive
execution allowlists).

\parab{Deployment and trust infrastructure}
Several infrastructure challenges must be addressed for production
deployment.  \emph{Kernel integrity assurance:} chaining the Agent Kernel's
trust anchor to a hardware root of trust via platform secure/measured boot,
signed kernel images, and runtime integrity monitoring, so that remote peers
can verify---before mutual attestation---that the mediation layer itself has
not been substituted or downgraded.  \emph{Cross-GAR federation:} as multiple
GARs emerge in a multi-stakeholder ecosystem, mutual recognition protocols
analogous to cross-certification in PKI are needed, together with formal
models of trust propagation across GAR boundaries.  \emph{Multi-agent
security composition:} when multiple \sys-protected agents interact, how do
their security invariants compose?  The intersection semantics (Principle~2)
extend naturally to pairwise A2A sessions, but multi-party composition
(e.g., a workflow spanning agents governed by different GARs with
incompatible taint lattices) raises open questions about invariant
preservation under heterogeneous trust models.

\parab{Ecosystem standards and advanced primitives}
The agent ecosystem needs shared abstractions---analogous to POSIX for
traditional OS services---for identity, memory provenance, and execution
governance.  Building on W3C Verifiable Credentials~\cite{w3c-vc-2.0} with
agent-specific semantics (capability boundaries, delegation chains,
revocation semantics) and standardizing the adapter interfaces would enable
interoperability across agent runtimes, governance platforms, and execution
sandboxes without fragmenting the security model.  On the cryptographic
front, three extensions are natural: \emph{privacy-preserving attestation}
via zero-knowledge proofs, enabling an agent to prove it possesses a required
capability without revealing its full identity; \emph{post-quantum readiness}
by supporting PQ signature algorithms (e.g., ML-DSA, SLH-DSA) as drop-in
replacements for Ed25519 in the AIC and attestation protocols; and
\emph{TEE-backed key custody} extending the tiered custody path
(\S\ref{sec:arch:identity}) with formal verification of the enclave
interface.

These four themes are interdependent: formal models inform what should be
empirically tested; deployment experience reveals which formal guarantees
matter most; standards encode what the ecosystem converges on.  Together they
define the path from a positioning paper to a deployed agent OS security
layer.  Section~\ref{sec:conclusion} summarizes our contributions toward this
agenda.

%% file: sections/conclusion.tex
\section{Conclusion}
\label{sec:conclusion}

Agents routinely cross trust boundaries in a single session; today's
orchestration frameworks, agent runtimes, governance platforms, and
execution sandboxes each cover part
of the problem, but none supplies a \emph{unified, mandatory} layer that
mediates identity, perception, memory, and execution end-to-end.
We argued that agents need an operating system, not only a stronger
harness, and that security should be the organizing principle of that
layer.

\sys is such an agent OS.  Four pillars---Identity, Perception,
Cognition, and Execution---enforce mediation at every boundary the agent
crosses, grounded in classical reference-monitor and
information-flow ideas and extended to semantic threats such as prompt
injection, memory poisoning, and unsafe tool use.  A unifying theme
runs through the design: security, when made structural rather than
policy-based, simultaneously enables richer agent capabilities.
Kernel-managed identity supports trustworthy collaboration across
agents and organizations; graduated perception lets agents safely
ingest richer input channels; taint-aware memory separates reliable
knowledge from contaminated content; and kernel-backed execution
mediation lets operators widen tool access without abandoning
non-bypassable guarantees.  \sys sits beneath the
broader harness ecosystem as a complementary OS layer rather than a
replacement for orchestration frameworks, agent runtimes, governance
platforms, or execution sandboxes.

As agents take on higher-stakes tasks, the same class of services that
traditional kernels provided for applications---strong identity,
mediated I/O, governed memory, and controlled execution---will be
indispensable at the semantic level.  We present \sys as a concrete
step toward that foundation and hope it informs further work on
agent-native operating systems.

%% file: bibliography/refs.bib
@misc{blocka2a,
  title   = {{BlockA2A}: Towards Secure and Verifiable Agent-to-Agent Interoperability},
  author  = {Zhenhua Zou and Zhuotao Liu and Lepeng Zhao and Qiuyang Zhan},
  year    = {2025},
  eprint  = {2508.01332},
  archivePrefix = {arXiv},
  primaryClass  = {cs.CR},
  url     = {https://arxiv.org/abs/2508.01332},
}

@article{jia2024acos,
  title   = {Agent Centric Operating System -- a Comprehensive Review and Outlook for Operating System},
  author  = {Shian Jia and Xinbo Wang and Mingli Song and Gang Chen},
  journal = {arXiv preprint arXiv:2411.17710},
  year    = {2024},
}

@article{mei2024aios,
  title   = {{AIOS}: {LLM} Agent Operating System},
  author  = {Kai Mei and Xi Zhu and Wujiang Xu and Wenyue Hua and Mingyu Jin and Zelong Li and Shuyuan Xu and Ruosong Ye and Yingqiang Ge and Yongfeng Zhang},
  journal = {arXiv preprint arXiv:2403.16971},
  year    = {2024},
}

@article{ge2023llmos,
  title   = {{LLM} as {OS}, Agents as Apps: Envisioning {AIOS}, Agents and the {AIOS}-Agent Ecosystem},
  author  = {Yingqiang Ge and Yujie Ren and Wenyue Hua and Shuyuan Xu and Juntao Tan and Yongfeng Zhang},
  journal = {arXiv preprint arXiv:2312.03815},
  year    = {2023},
}

@article{wu2024oscopilot,
  title   = {{OS-Copilot}: Towards Generalist Computer Agents with Self-Improvement},
  author  = {Zhiyong Wu and Chengcheng Han and Zichen Ding and Zhenmin Weng and Zhoumianze Liu and Shunyu Yao and Tao Yu and Lingpeng Kong},
  journal = {arXiv preprint arXiv:2402.07456},
  year    = {2024},
}

@article{wang2024openhands,
  title   = {{OpenHands}: An Open Platform for {AI} Software Developers as Generalist Agents},
  author  = {Xingyao Wang and Boxuan Li and Yufan Song and Frank F. Xu and Xiangru Tang and others},
  journal = {arXiv preprint arXiv:2407.16741},
  year    = {2024},
}

@article{agashe2024agents,
  title   = {Agent {S}: An Open Agentic Framework that Uses Computers Like a Human},
  author  = {Saaket Agashe and Jiuzhou Han and Shuyu Gan and Jiachen Yang and Ang Li and Xin Eric Wang},
  journal = {arXiv preprint arXiv:2410.08164},
  year    = {2024},
}

@article{bonatti2024windowsarena,
  title   = {Windows Agent Arena: Evaluating Multi-Modal {OS} Agents at Scale},
  author  = {Rogerio Bonatti and Dan Zhao and Francesco Bonacci and Dillon Dupont and Sara Abdali and others},
  journal = {arXiv preprint arXiv:2409.08264},
  year    = {2024},
}

@article{wu2024osatlas,
  title   = {{OS-ATLAS}: A Foundation Action Model for Generalist {GUI} Agents},
  author  = {Zhiyong Wu and Zhenyu Wu and Fangzhi Xu and Yian Wang and Qiushi Sun and others},
  journal = {arXiv preprint arXiv:2410.23218},
  year    = {2024},
}

@article{hu2025osagents,
  title   = {{OS} Agents: A Survey on {MLLM}-based Agents for General Computing Devices Use},
  author  = {Xueyu Hu and Tao Xiong and Biao Yi and Zishu Wei and Ruixuan Xiao and others},
  journal = {arXiv preprint arXiv:2508.04482},
  year    = {2025},
}

@article{li2025memos,
  title   = {{MemOS}: A Memory {OS} for {AI} System},
  author  = {Zhiyu Li and Chenyang Xi and Chunyu Li and Ding Chen and Boyu Chen and others},
  journal = {arXiv preprint arXiv:2507.03724},
  year    = {2025},
}

@article{kuntz2025osharm,
  title   = {{OS-Harm}: A Benchmark for Measuring Safety of Computer Use Agents},
  author  = {Thomas Kuntz and Agatha Duzan and Hao Zhao and Francesco Croce and Zico Kolter and Nicolas Flammarion and Maksym Andriushchenko},
  journal = {arXiv preprint arXiv:2506.14866},
  year    = {2025},
}

@article{basu2026toolreceipts,
  title   = {Tool Receipts, Not Zero-Knowledge Proofs: Practical Hallucination Detection for {AI} Agents},
  author  = {Abhinaba Basu},
  journal = {arXiv preprint arXiv:2603.10060},
  year    = {2026},
}

@article{ye2025mobileagentv3,
  title   = {Mobile-Agent-v3: Fundamental Agents for {GUI} Automation},
  author  = {Jiabo Ye and Xi Zhang and Haiyang Xu and Haowei Liu and Junyang Wang and others},
  journal = {arXiv preprint arXiv:2508.15144},
  year    = {2025},
}

@article{fourney2024magneticone,
  title   = {Magentic-One: A Generalist Multi-Agent System for Solving Complex Tasks},
  author  = {Adam Fourney and Gagan Bansal and Hussein Mozannar and Cheng Tan and Eduardo Salinas and others},
  journal = {arXiv preprint arXiv:2411.04468},
  year    = {2024},
}

@article{tang2025autoagent,
  title   = {{AutoAgent}: A Fully-Automated and Zero-Code Framework for {LLM} Agents},
  author  = {Jiabin Tang and Tianyu Fan and Chao Huang},
  journal = {arXiv preprint arXiv:2502.05957},
  year    = {2025},
}

@article{tang2025guisurvey,
  title   = {A Survey on ({M}){LLM}-Based {GUI} Agents},
  author  = {Fei Tang and Haolei Xu and Hang Zhang and Siqi Chen and Xingyu Wu and others},
  journal = {arXiv preprint arXiv:2504.13865},
  year    = {2025},
}

@article{yang2025agentprotocols,
  title   = {A Survey of {AI} Agent Protocols},
  author  = {Yingxuan Yang and Huacan Chai and Yuanyi Song and Siyuan Qi and Muning Wen and others},
  journal = {arXiv preprint arXiv:2504.16736},
  year    = {2025},
}

@article{dong2024agentops,
  title   = {{AgentOps}: Enabling Observability of {LLM} Agents},
  author  = {Liming Dong and Qinghua Lu and Liming Zhu},
  journal = {arXiv preprint arXiv:2411.05285},
  year    = {2024},
}

@article{wei2025agentxpu,
  title   = {Agent.xpu: Efficient Scheduling of Agentic {LLM} Workloads on Heterogeneous {SoC}},
  author  = {Xinming Wei and Jiahao Zhang and Haoran Li and Jiayu Chen and others},
  journal = {arXiv preprint arXiv:2506.24045},
  year    = {2025},
}

@article{wang2026openclaw,
  title   = {From Assistant to Double Agent: Formalizing and Benchmarking Attacks on {OpenClaw} for Personalized Local {AI} Agent},
  author  = {Yuhang Wang and Feiming Xu and Zheng Lin and Guangyu He and others},
  journal = {arXiv preprint arXiv:2602.08412},
  year    = {2026},
}

@misc{openclaw2026github,
  author       = {{OpenClaw}},
  title        = {{OpenClaw}: Personal {AI} assistant framework},
  howpublished = {\url{https://github.com/openclaw/openclaw}},
  year         = {2026},
  note         = {Open-source repository; accessed May 2026},
}

@misc{nous2026hermesagent,
  author       = {{Nous Research}},
  title        = {{Hermes Agent}: Open-source agent harness with autonomous skills and scheduling},
  howpublished = {\url{https://github.com/NousResearch/hermes-agent}},
  year         = {2026},
  note         = {Accessed May 2026},
}

@article{zhang2024aiinsideos,
  title   = {Integrating Artificial Intelligence into Operating Systems: A Survey on Techniques, Applications, and Future Directions},
  author  = {Yifan Zhang and Xinkui Zhao and Ziying Li and Guanjie Cheng and Jianwei Yin and others},
  journal = {arXiv preprint arXiv:2407.14567},
  year    = {2024},
}

@inproceedings{kang2025memoryos,
  title     = {Memory {OS} of {AI} Agent},
  author    = {Jiazheng Kang and Mingming Ji and Zhe Zhao and Ting Bai},
  booktitle = {Proceedings of EMNLP},
  pages     = {25961--25970},
  year      = {2025},
  doi       = {10.18653/v1/2025.emnlp-main.1318},
}

@inproceedings{zhang2025ufo,
  title     = {{UFO}: A {UI}-Focused Agent for {W}indows {OS} Interaction},
  author    = {Chaoyun Zhang and Liqun Li and Shilin He and Xu Zhang and Bo Qiao and others},
  booktitle = {Proceedings of NAACL},
  year      = {2025},
}

@inproceedings{jia2025agentstore,
  title     = {{AgentStore}: Scalable Integration of Heterogeneous Agents as Specialized Generalist Computer Assistant},
  author    = {Chengyou Jia and Minnan Luo and Zhuohang Dang and Qiushi Sun and others},
  booktitle = {Findings of ACL},
  pages     = {8908--8934},
  year      = {2025},
  url       = {https://aclanthology.org/2025.findings-acl.466/},
}

@incollection{zhuo2025kaos,
  title     = {{KAOS}: Large Model Multi-agent Operating System},
  author    = {Zhao Zhuo and Rongzhen Li and Kai Liu and Huhai Zou and others},
  booktitle = {CCIS (CCKS-IJCKG 2024)},
  publisher = {Springer},
  year      = {2025},
}

@misc{autogpt2025,
  title        = {{AutoGPT}: The Vision of Accessible {AI} for Everyone},
  author       = {{Significant Gravitas}},
  year         = {2025},
  howpublished = {\url{https://github.com/Significant-Gravitas/AutoGPT}},
  note         = {183\,K GitHub stars as of Apr 2026},
}

@misc{langchain2025,
  title        = {{LangChain}: The Agent Engineering Platform},
  author       = {{LangChain, Inc.}},
  year         = {2025},
  howpublished = {\url{https://github.com/langchain-ai/langchain}},
  note         = {137\,K GitHub stars as of May 2026},
}

@misc{langsmith2025,
  title        = {{LangSmith}: {AI} Agent \& {LLM} Observability Platform},
  author       = {{LangChain, Inc.}},
  year         = {2025},
  howpublished = {\url{https://www.langchain.com/langsmith/}},
  note         = {Framework-agnostic tracing, evaluation, and monitoring
                  for LLM applications and AI agents},
}

@misc{langgraph2025,
  title        = {{LangGraph}: Build Resilient Language Agents as Graphs},
  author       = {{LangChain, Inc.}},
  year         = {2025},
  howpublished = {\url{https://github.com/langchain-ai/langgraph}},
  note         = {32\,K GitHub stars as of May 2026},
}

@misc{autogen2024,
  title        = {{AutoGen}: A Programming Framework for Agentic {AI}},
  author       = {{Microsoft}},
  year         = {2024},
  howpublished = {\url{https://github.com/microsoft/autogen}},
  note         = {57\,K GitHub stars as of Apr 2026; README states maintenance mode (no new features) with migration to Microsoft Agent Framework; see repository \url{https://github.com/microsoft/autogen} and \url{https://learn.microsoft.com/en-us/agent-framework/migration-guide/from-autogen/}},
}

@misc{crewai2025,
  title        = {{CrewAI}: Framework for Orchestrating Role-Playing, Autonomous {AI} Agents},
  author       = {{CrewAI, Inc.}},
  year         = {2025},
  howpublished = {\url{https://github.com/crewAIInc/crewAI}},
  note         = {51\,K GitHub stars as of May 2026},
}

@misc{semantickernel2025,
  title        = {{Semantic Kernel}: An {SDK} for Integrating {LLM} Technology into Apps},
  author       = {{Microsoft}},
  year         = {2025},
  howpublished = {\url{https://github.com/microsoft/semantic-kernel}},
  note         = {28\,K GitHub stars as of Apr 2026},
}

@misc{smolagents2025,
  title        = {smolagents: A Barebones Library for Agents That Think in Code},
  author       = {{Hugging Face}},
  year         = {2025},
  howpublished = {\url{https://github.com/huggingface/smolagents}},
  note         = {27.3\,K GitHub stars as of May 2026},
}

@article{metagpt2024,
  title   = {{MetaGPT}: Meta Programming for a Multi-Agent Collaborative Framework},
  author  = {Sirui Hong and Mingchen Zhuge and Jonathan Chen and Xiawu Zheng and Yuheng Cheng and others},
  journal = {arXiv preprint arXiv:2308.00352},
  year    = {2024},
}

@inproceedings{yang2024sweagent,
  title     = {{SWE-agent}: Agent-Computer Interfaces Enable Automated Software Engineering},
  author    = {John Yang and Carlos E. Jimenez and Alexander Wettig and Kilian Lieret and Shunyu Yao and Karthik Narasimhan and Ofir Press},
  booktitle = {NeurIPS},
  year      = {2024},
}

@inproceedings{xie2024osworld,
  title     = {{OSWorld}: Benchmarking Multimodal Agents for Open-Ended Tasks in Real Computer Environments},
  author    = {Tianbao Xie and Danyang Zhang and Jixuan Chen and Xiaochuan Li and Siheng Zhao and others},
  booktitle = {NeurIPS},
  year      = {2024},
}

@misc{south2025identitymanagement,
  title   = {Identity Management for Agentic {AI}: The New Frontier of Authorization, Authentication, and Security for an {AI} Agent World},
  author  = {Tobin South and Subramanya Nagabhushanaradhya and others},
  year    = {2025},
  eprint  = {2510.25819},
  archivePrefix = {arXiv},
  primaryClass  = {cs.CR},
}

@inproceedings{greshake2023indirect,
  title     = {Not What You've Signed up For: Compromising Real-World {LLM}-Integrated Applications with Indirect Prompt Injection},
  author    = {Kai Greshake and Sahar Abdelnabi and Shailesh Mishra and Christoph Endres and Thorsten Holz and Mario Fritz},
  booktitle = {Proceedings of the 16th ACM Workshop on Artificial Intelligence and Security},
  pages     = {79--90},
  year      = {2023},
}

@inproceedings{shen2024jailbreak,
  title     = {``Do Anything Now'': Characterizing and Evaluating In-the-Wild Jailbreak Prompts on Large Language Models},
  author    = {Xinyue Shen and Zeyuan Chen and Michael Backes and Yun Shen and Yang Zhang},
  booktitle = {Proceedings of ACM CCS},
  pages     = {1671--1685},
  year      = {2024},
}

@article{denning1976lattice,
  title   = {A Lattice Model of Secure Information Flow},
  author  = {Dorothy E. Denning},
  journal = {Communications of the ACM},
  volume  = {19},
  number  = {5},
  pages   = {236--243},
  year    = {1976},
}

@article{bell1973secure,
  title   = {Secure Computer Systems: Mathematical Foundations},
  author  = {David Elliott Bell and Leonard J. LaPadula},
  journal = {MITRE Technical Report},
  number  = {MTR-2547},
  year    = {1973},
}

@article{biba1977integrity,
  title   = {Integrity Considerations for Secure Computer Systems},
  author  = {Kenneth J. Biba},
  journal = {MITRE Technical Report},
  number  = {MTR-3153},
  year    = {1977},
}

@misc{sgxexplained,
  title        = {Intel {SGX} Explained},
  author       = {Victor Costan and Srinivas Devadas},
  howpublished = {Cryptology ePrint Archive, Paper 2016/086},
  year         = {2016},
}

@inproceedings{ngabonziza2016trustzone,
  title     = {{TrustZone} Explained: Architectural Features and Use Cases},
  author    = {Bernard Ngabonziza and Daniel Martin and Anna Bailey and Haehyun Cho and Sarah Martin},
  booktitle = {IEEE CIC},
  pages     = {445--451},
  year      = {2016},
}

@techreport{w3c-vc-2.0,
  title       = {Verifiable Credentials Data Model v2.0},
  author      = {Manu Sporny and others},
  institution = {W3C},
  type        = {W3C Recommendation},
  year        = {2025},
  url         = {https://www.w3.org/TR/2025/REC-vc-data-model-2.0-20250515/},
}

@misc{anthropic2024mcp,
  title        = {Model Context Protocol ({MCP}) Specification},
  author       = {{Anthropic}},
  year         = {2024},
  howpublished = {\url{https://modelcontextprotocol.io/}},
}

@misc{google2025a2a,
  title        = {Agent-to-Agent ({A2A}) Protocol},
  author       = {{Google}},
  year         = {2025},
  howpublished = {\url{https://a2aproject.github.io/A2A/latest/}},
}

@inproceedings{yao2023react,
  title     = {{ReAct}: Synergizing Reasoning and Acting in Language Models},
  author    = {Shunyu Yao and Jeffrey Zhao and Dian Yu and Nan Du and Izhak Shafran and Karthik Narasimhan and Yuan Cao},
  booktitle = {ICLR},
  year      = {2023},
}

@misc{microsoft2025governance,
  title        = {Agent Governance Toolkit: Runtime Governance for {AI} Agents},
  author       = {{Microsoft}},
  year         = {2026},
  howpublished = {\url{https://github.com/microsoft/agent-governance-toolkit}},
  note         = {Modular platform comprising Agent~OS (policy engine),
                  AgentMesh (zero-trust SPIFFE/SVID identity), Agent Hypervisor
                  (execution rings), Agent~SRE, compliance verification,
                  and shadow-AI discovery; 13\,000+ tests; Python /
                  TypeScript / .NET / Rust / Go SDKs; 1.6\,K GitHub stars as of May 2026},
}

@misc{agentmesh,
  title        = {{AgentMesh}: Production-Grade Trust Layer for Multi-Agent Systems},
  author       = {{Microsoft}},
  year         = {2026},
  howpublished = {\url{https://github.com/microsoft/agent-governance-toolkit/tree/main/agent-governance-python/agent-mesh}},
  note         = {Part of the Agent Governance Toolkit; SPIFFE/SVID-based identity, policy engine, A2A/MCP/IATP protocol bridge},
}

@article{hdp,
  title   = {{HDP}: A Lightweight Cryptographic Protocol for Human Delegation Provenance in Agentic {AI} Systems},
  author  = {Asiri Dalugoda},
  journal = {arXiv preprint arXiv:2604.04522},
  year    = {2026},
  url     = {https://arxiv.org/abs/2604.04522},
}

@article{zt-agentic-ai,
  title   = {A Novel Zero-Trust Identity Framework for Agentic {AI}: Decentralized Authentication and Fine-Grained Access Control},
  author  = {Ken Huang and Vineeth Sai Narajala and John Yeoh and Jason Ross and others},
  journal = {arXiv preprint arXiv:2505.19301},
  year    = {2025},
}

@article{anp-whitepaper,
  title   = {Agent Network Protocol Technical White Paper},
  author  = {Gao Chang and Eddie Lin and Chongyuan Yuan and Ran Cai and Bin Chen and Xinlin Xie and others},
  journal = {arXiv preprint arXiv:2508.00007},
  year    = {2025},
}

@article{agent-osi,
  title   = {{Agent-OSI}: A Layered Protocol Stack Toward a Decentralized Internet of Agents},
  author  = {Weijie Xu and Tianle Wang and Yuxuan Xia and Sheng Zhang and Soung Chang Liew},
  journal = {arXiv preprint arXiv:2602.13795},
  year    = {2026},
}

@inproceedings{anderson1972planning,
  title     = {Computer Security Technology Planning Study},
  author    = {James P. Anderson},
  booktitle = {Technical Report ESD-TR-73-51, USAF Electronic Systems Division},
  year      = {1972},
}

@article{saltzer1975protection,
  title   = {The Protection of Information in Computer Systems},
  author  = {Jerome H. Saltzer and Michael D. Schroeder},
  journal = {Proceedings of the IEEE},
  volume  = {63},
  number  = {9},
  pages   = {1278--1308},
  year    = {1975},
}

@misc{openfang2026,
  title        = {{OpenFang}: Open-source Agent Operating System},
  author       = {{RightNow AI}},
  year         = {2026},
  howpublished = {\url{https://github.com/RightNow-AI/openfang}},
  note         = {17.5\,K GitHub stars as of May 2026; Rust; WASM sandbox,
                  Merkle audit chain, Ed25519 signed manifests,
                  prompt-injection scanning, MCP/A2A support},
}

@misc{nono2026,
  title        = {nono: A Capability-Based Sandbox for {AI} Agents},
  author       = {{always-further}},
  year         = {2026},
  howpublished = {\url{https://github.com/always-further/nono}},
  note         = {2.4\,K GitHub stars as of May 2026; Landlock/Seatbelt
                  kernel sandboxing, Sigstore signing of instruction files,
                  credential proxy injection, network allowlists},
}

@misc{smythos2026,
  title        = {{SmythOS SRE}: Cloud-Native Runtime for Agentic {AI}},
  author       = {{SmythOS}},
  year         = {2026},
  howpublished = {\url{https://github.com/SmythOS/sre}},
  note         = {1.3\,K GitHub stars as of May 2026; TypeScript;
                  kernel + SDK + CLI monorepo with unified resource
                  abstractions and ACL-based access control},
}

@misc{letta2026,
  title        = {Letta: The Platform for Building Stateful Agents with Advanced Memory},
  author       = {{Letta AI}},
  year         = {2026},
  howpublished = {\url{https://github.com/letta-ai/letta}},
  note         = {22.7\,K GitHub stars as of May 2026; formerly MemGPT;
                  layered memory management and context control},
}

@misc{anthropic-sandbox2026,
  title        = {Sandbox Runtime: Lightweight {OS}-Level Sandboxing Without Containers},
  author       = {{Anthropic}},
  year         = {2026},
  howpublished = {\url{https://github.com/anthropic-experimental/sandbox-runtime}},
  note         = {4.1\,K GitHub stars as of May 2026; process-level
                  filesystem and network restrictions at the OS level},
}

@misc{e2b2026,
  title        = {{E2B}: Open-source Secure Environments for {AI} Agents},
  author       = {{E2B}},
  year         = {2026},
  howpublished = {\url{https://github.com/e2b-dev/E2B}},
  note         = {12.2\,K GitHub stars as of May 2026; cloud-side
                  isolated sandboxes for AI-generated code execution},
}

@misc{ms-agent-framework2026,
  title        = {Microsoft Agent Framework},
  author       = {{Microsoft}},
  year         = {2026},
  howpublished = {\url{https://github.com/microsoft/agent-framework}},
  note         = {10.5\,K GitHub stars as of May 2026; successor to
                  AutoGen/Semantic Kernel; Python/.NET multi-agent SDK},
}

@misc{anysphere2026cursor,
  author       = {{Anysphere}},
  title        = {{Cursor}: {AI}-native editor, tab completion, and coding agents},
  howpublished = {\url{https://cursor.com/}},
  year         = {2026},
  note         = {Commercial product; accessed May 2026},
}

@misc{anthropic2026claudecode,
  author       = {{Anthropic}},
  title        = {{Claude Code}: Agentic coding tool for terminal and {IDE}},
  howpublished = {\url{https://claude.com/product/claude-code}},
  year         = {2026},
  note         = {Commercial product; accessed May 2026},
}

@misc{github2026copilot,
  author       = {{GitHub}},
  title        = {{GitHub Copilot} documentation},
  howpublished = {\url{https://docs.github.com/copilot}},
  year         = {2026},
  note         = {Official documentation hub; accessed May 2026},
}

@misc{github2026openaicodex,
  author       = {{GitHub}},
  title        = {OpenAI Codex coding agent in {GitHub Copilot}},
  howpublished = {\url{https://docs.github.com/en/copilot/concepts/agents/openai-codex}},
  year         = {2026},
  note         = {Coding agent documented under Copilot; accessed May 2026},
}
